\documentclass[universe,review,accept,pdftex,moreauthors]{Definitions/mdpi} 

\firstpage{1} 
\pubvolume{1}
\issuenum{1}
\articlenumber{0}
\pubyear{2025}
\copyrightyear{2025}
\externaleditor{Name}
\datereceived{27 December 2024} 
\daterevised{24 February 2025} 
\dateaccepted{25 February 2025} 
\datepublished{ } 
\hreflink{https://doi.org/} 
\usepackage{latexsym,amsmath,amssymb}
\usepackage{graphics,graphicx}
\usepackage{natbib}
\usepackage{rotating}
\usepackage{multibib}

\newcommand{\nh}{N${\rm _H}$}

\newcommand{\rxte}{{\it RXTE}}
\newcommand{\asca}{{\it ASCA}}
\newcommand{\rosat}{{\it ROSAT}}
\newcommand{\Einstein}{{\it EINSTEIN}}

\newcommand{\exo}{{\it EXOSAT}}
\newcommand{\erosita}{{\it eROSITA}}
\newcommand{\suz}{{\it Suzaku}}
\newcommand{\xmm}{{\it XMM-Newton}}
\newcommand{\sax}{{\it BeppoSAX}}
\newcommand{\swi}{{\it Swift}}
\newcommand{\cha}{{\it Chandra}}
\newcommand{\nustar}{\textit{NuSTAR}}
\newcommand{\nicer}{\textit{NICER}}
\newcommand{\integral}{\textit{INTEGRAL}}
\newcommand{\hst}{{\it HST}}
\newcommand{\vla}{{\it VLA}}
\newcommand{\msun}{${\rm M}_{\odot}$}
\newcommand{\rsun}{$R_{\odot}$}

\newcommand{\lumcgs}{erg~s$^{-1}$}

\def\aj{AJ}
\def\apj{ApJ}

\def\aap{A\&A}
\def\mnras{MNRAS}
\def\apjs{ApJS}

\def\nat{Nature}
\def\jaavso{JAAVSO}

\def\apjl{ApJLett}

\def\pasp{PASP}
\def\pasj{PASJ}
\def\nar{NewAR}
\def\araa{ARA\&A}
\def\actaa{Ac. Ast.}

\Title{Broad Observational Perspectives Achieved by the Accreting White Dwarf Sciences in the {\it \textbf{XMM-Newton}} and {\it \textbf{Chandra}} Eras}

\TitleCitation{Accreting WD Sciences in the {\it XMM-Newton} and {\it Chandra} Eras}

\Author{ \c{S}\"olen {Balman} 
 $^{1,2,}$*\orcidA{}, Marina Orio $^{3,4}$\orcidB{} and Gerardo J. M. Luna $^{5,6}$\orcidC{}}

\AuthorNames{\c{S}\"olen Balman, Marina Orio and Gerardo J. M. Luna}

\AuthorCitation{Balman, \c{S}.; Orio, M.; Luna, G.J.M.}

\address{%
$^{1}$ \quad Department of Astronomy and Space Sciences, Faculty of Science, Istanbul University, Beyazit, \mbox{34119 Istanbul, Turkey}\\
$^{2}$ \quad Faculty of Engineering and Natural Sciences, Kadir Has University, Cibali, 34083 Istanbul, Turkey\\
{ $^{3}$ \quad Department of Astronomy, University of Wisconsin, 475 N. Charter Str., Madison, WI 53706, USA; orio@astro.wisc.edu\\
$^{4}$ \quad INAF-Osservatorio Astronomico di Padova, Vicolo Osservatorio, 5, 1-35122 Padova, Italy\\}
{ $^{5}$ \quad Secretar{\'i}a de Investigaci{\'o}n, {Universidad} Nacional de Hurlingham (UNAHUR), Av. Gdor. Vergara 2222, \linebreak Villa Tesei, {Buenos Aires, B1688}, 
 Argentina; juan.luna@unahur.edu.ar\\}
{ $^{6}$ \quad Consejo Nacional de Investigaciones Cient{\'i}ficas y T{\'e}cnicas (CONICET), Argentina}}
\corres{Correspondence: solen.balman@istanbul.edu.tr, solen.balman@gmail.com; Tel.: +90-533-4657232)}

\abstract{Accreting white dwarf binaries (AWDs) comprise cataclysmic variables (CVs), symbiotics, AM CVns, and other related systems that host a primary white dwarf (WD) accreting from a main sequence or evolved companion star. AWDs are a product of close binary evolution; thus, they are important for understanding the evolution and population of X-ray binaries in the Milky Way and other galaxies. AWDs are essential for studying astrophysical plasmas under different conditions along with accretion physics and processes, transient events, matter ejection and outflows, compact binary evolution, mergers, angular momentum loss mechanisms, and nuclear processes leading to explosions. AWDs are also closely related to other  objects in the late stages of stellar evolution, with other accreting objects in compact binaries, and even share common phenomena with young stellar objects, active galactic nuclei, quasars, and supernova remnants.  As  X-ray astronomy came to a climax with the start of the \cha\ and \xmm\ missions owing to their unprecedented instrumentation,  new excellent imaging capabilities, good time resolution, and X-ray grating technologies allowed immense advancement in many aspects of astronomy and astrophysics. In this review, we lay out a panorama of developments on the study of AWDs that have been accomplished and have been made possible by these two observatories; we summarize the key observational achievements and the challenges ahead.}


\begin{document}

\section{Introduction}
\subsection{A Brief History of X-ray Astronomy}
X-ray astronomy is a product of the Space Age. X-rays from outside the Earth were discovered in 1949, when radiation detectors aboard rockets above the atmosphere measured X-rays coming from the Sun. Following this, it took more than a decade and improved instrumentation to discover X-rays  from sources beyond the solar system. Bruno Rossi inferred that several astrophysical sources must be sources of X-rays, and while the motivation for scanning the sky at X-ray energies was ``justified'' as observing the Moon, in 1962, with an Aereobee rocket, {{Riccardo} Giacconi (and collaborators, Gursky, Paolini, and Rossi}) discovered Scorpius X-1, the first confirmed astrophysical source of X-rays (a neutron star X-ray binary).

This landmark discovery paved the way for the National Aeronautics and Space Administration (NASA)'s {\it {Uhuru}} X-ray scanning satellite, which was sent for its mission at the end of 1970,  equipped with two proportional counters, with a viewing pipe to locate the X-ray sources. {\it Uhuru} allowed {{discoveries}/detections} such as evidence of black holes and superdense neutron stars pulling matter from companion stars, and hot gas in galaxy clusters. Using an all-sky survey, it discovered 339 X-ray sources. The first large focusing X-ray telescope was the Apollo Telescope onboard the {\it Skylab} in the early 1970s. This pioneering telescope laid the groundwork for the development of the \Einstein\ X-ray Observatory by NASA.
{{In 1951}, a German physicist Hans Wolter discovered a mirror configuration to produce a focusing X-ray telescope. It consisted of a paraboloid and hyperboloid mirror mounted confocally and coaxially (nested-mirror technology for focusing X-rays). Such Wolter telescopes were first used on the {\it Skylab} to investigate the corona of the Sun. This was followed in 1978 by NASA's Einstein Observatory, and in 1983 by European Space Agency (ESA)'s EXOSAT, both equipped with Wolter-type telescopes. This technology was improved and used in several other (focusing) X-ray imaging  telescopes in the later eras (e.g., \cha\ and \xmm)}.

The \Einstein\ Observatory, launched in 1978, was the first large imaging X-ray telescope, with  actual X-ray mirrors. It accomplished many milestones, like imaging the consequence of shock waves in supernova remnants, and the hot gas in galaxies and clusters of galaxies.  \Einstein\ also accurately located over 7000 X-ray sources and found new ways to study the dark matter surrounding galaxies. Another historical mission to note is the \exo\ (The European X-ray Observatory SATellite), which was operational between May 1983 and  April 1986 and observed 1780 astronomical objects. The payload instruments were used to produce spectra, images, and light curves in various EUV and X-ray energy bands. The Roentgensatellite, or \rosat, a collaboration between Germany, the United Kingdom, and the United States, carried an even larger X-ray telescope into space in 1990, leading to the discovery of more than 60,000 sources and has been especially valuable for investigating the multimillion-degree gas present in the upper atmospheres of many stars.   Its unprecedented sensitivity in the soft X-ray band of 0.1--2.4 keV  allowed many soft/supersoft X-ray sources to be detected, and related phenomena to be studied (e.g., stellar outbursts). The \rosat\ all-sky survey and its catalogs are still commonly used today.

\asca\ (Advanced Satellite for Cosmology and Astrophysics) launched in 1993, a joint mission by Japan and the United States, {was the first X-ray satellite to use CCDs (Charged Coupled Devices) as detectors, leading to a huge improvement in spectral resolution.}  It was especially crafted to study the spectra of hot gases in the X-ray range and reveal the chemical composition within the hot gas. \asca\ ceased operations in July of 2000. Contemporary to \asca, the NASA mission {\it Rossi X-ray Timing Explorer (RXTE)} was launched in December 1995. \rxte\ did not have focusing X-ray mirrors, but with a wide energy range of 2--150 keV, high sensitivity, and a unique capability to measure the rapid time variability of X-ray sources, it has made valuable contributions to our understanding of neutron stars, black holes, compact X-ray binaries, and their transient nature. The \sax\ satellite of the Italian Space Agency with the participation of the Netherlands Agency for Aerospace Programs and ESA was launched  in  April 1996, and operated until April 2002. It was the first X-ray mission that covered three decades of energy---from 0.1 to 300 keV---with moderate imaging capability. It proved to be exceptionally useful in discovering and imaging X-ray sources associated with Gamma-ray bursts, determining their positions with unprecedented precision, and monitoring the X-ray afterglow.  

The advancement of X-ray astronomy and X-ray telescopes came to a turning point with the next generation of all-purpose facilities. The \cha\ X-ray Observatory obtained initial funding in 1977 as the {\it Advanced X-ray Astrophysics Facility (AXAF)}. The project was restructured in 1992 and came to completion in 1999, with the current name.  {The X-ray telescope was equipped with unprecedented imaging capabilities (highest spatial resolution achieved in the X-rays) along with high sensitivity and good spectral resolution quality compared to its predecessors. A completely new technology and a new window to X-ray astronomy was introduced by the X-ray transmission gratings as part of the payload yielding the highest spectral resolution achieved at that time.}
Concomitant to the  development of the \cha\ X-ray observatory, ESA studied a  multi-purpose, high-throughput complementary X-ray facility.  The {\it X-ray Multi-Mirror Astronomy Mission, later named \xmm}, was proposed to ESA in 1982 as a mission optimized for high collecting power to accommodate maximum spectral sensitivity. It was approved in 1989, and the instrumentation and hardware development started. {The mission's imaging capability was modest compared to \cha, but had high-throughput CCD cameras with a huge collecting area due to its three co-aligned X-ray mirrors, and the payload was also equipped with another new technology; reflection gratings provided very high spectral resolution in a narrower band compared to \cha. The advantage of \xmm\ was that all its instruments operated simultaneously. This ESA satellite was developed and built by Airbus and an industrial consortium of 45 European companies and 1 US company.}

After the launch of these two X-ray observatories, the X-ray astronomy came to a climax, marking the start of a new era. {Contemporary to these missions, a number of more recent X-ray observatories (see \url{https://heasarc.gsfc.nasa.gov}, for the mission web pages) with limited scopes, like \integral, \swi, \nicer, \nustar, and {\it IXPE} (first X-ray polarization mission), possessing extensive timing capabilities along with moderate-to-good spectral capabilities, provided a way to explore the X-ray sky while other all-purpose observatories were being planned. Such smaller missions  complemented our knowledge of astrophysics in the X-rays; they often served as pathfinders and monitors of variable and transient sources. The newest of such missions (as of writing of this review) is the X-ray Imaging and Spectroscopy Mission, {\it XRISM}, with a payload designed to provide the highest spectral resolution to date in the X-rays ($\sim$7 eV in 0.4--12 keV; it includes an X-ray calorimeter spectrometer) which started observations in 2024.}

In this review, 25 years after their launch, we outline the broad observational perspectives achieved in the accreting white dwarf sciences in the era of the two multi-purpose X-ray observatories, \cha\ and \xmm.

\subsection{The \xmm\ Mission}

The \xmm\ mission name stands for {\it X-ray Multi-Mirror Mission} and honors Sir Isaac Newton. In a broad sense, it observes and studies the X-ray sources in the universe looking deeply into external galaxies, studying stars at all evolutionary stages, investigating black holes  of all sizes and their environments, following up on explosive events like gamma-ray bursts (GRBs), supernovae, novae,  and other energetic astronomical persistent sources or transient phenomena, exploring the formation of the universe and how matter behaves under extreme conditions. \xmm\ is an European Space Agency (ESA) science mission with instruments and contributions funded by ESA member states including a contribution by the National Aeronautics and Space Administration (NASA). It was launched on December 10, 1999, on an Ariane-5 rocket from ESA's spaceport in French Guiana. It was designed to be fully functional for 10 years, but presently, it is approved until  December 2026, with the possibility of extensions. It has a 48-hour elliptical orbit from 7000 km to 114,000 km altitude.  \xmm\ is laying the groundwork for future missions,  like Advanced Telescope for High-Energy Astrophysics, {\it NewAthena}, and Laser Interferometer Space Antenna, {\it LISA}, exploring both the hot and energetic universe and the powerful electromagnetic counterparts of gravitational waves. \xmm\ is the biggest science satellite built in Europe, weighing 3.8 tons, with a total length of 16 m with its solar~arrays. 

There are six scientific instruments onboard \xmm: (1) the European Photon Imaging Cameras (EPIC: pn, MOS1,2), (2) the Reflection Grating Spectrometers (RGS1,2), and (3)  the Optical Monitor (OM) \citep{2001Jansen}. \xmm\ has three 1500 cm$^{2}$ X-ray telescopes, each equipped with  EPIC cameras, two of which have metal oxide semiconductor (MOS) CCDs~\citep{2001Turner} and the last one uses pn CCDs \citep{2001Struder} for data recording. The EPIC cameras collect data in the 0.2--12 keV range with a field of view (FOV) of $\sim$30 arc min diameter (PSF (FWHM) $\sim$6 arcsec), at a time resolution of about 70 ms (7 micro-s for burst and 0.03 ms for timing mode) {with a spectral resolution, e.g., E/$\Delta$E$\sim$20--50 at 6.5 keV}.  Around half of the incoming X-rays are diverted by reflection grating arrays RGS 1,2 \citep{2001denHerder}, which provide high-resolution ($\lambda$/$\Delta\lambda$$\sim$100--800) X-ray spectroscopy in the 0.33--2.5 keV energy band. RGS 1 and 2  provide data primarily for atomic spectroscopy and line diagnosis.
The optical monitor is a 2-meter-long telescope with a 30 cm diameter with imaging capability in three broadband ultraviolet filters and three optical filters (1800--6000 \AA), providing observations simultaneously with those in X-rays obtained by the EPIC cameras and the RGS, with a minimum time resolution of 0.5 s \citep{2001Mason}. 

\subsection{The \cha\ X-ray Observatory}

The \cha\ X-ray Observatory (CXO) is a flagship-class space telescope launched with the space shuttle Columbia by NASA on 23 July, 1999 \citep{2002Weisskopf}. \cha\ is  sensitive to X-ray sources  100 times fainter compared to its pre-2000 X-ray telescopes and has an unprecedented spatial resolution of 0.49 arcsec per pixel {(PSF FWHM $<$ 0.5 arcsec on-axis)}; the on-axis resolution will not be replaced in the future missions to come.  The elliptical orbit takes the spacecraft more than a third of the way to the moon before returning to its closest point to Earth at 16,000 km with a period of 64 h and 18 min. 
Chandra explores the hot turbulent regions in space with images 25 times sharper than previous X-ray images. 
\cha's improved sensitivity and exceptional resolution allows for detailed studies of black holes, supernovas, and dark matter,  increasing our understanding of the origin, evolution, and destiny of the universe together with studies on stellar structures, transient sources, and compact binary stars.

\cha\ uses four pairs of nested mirrors, together with their support structure, called the High-Resolution Mirror Assembly (HRMA), which directs the incoming X-rays to two focal plane cameras and two sets of transmission gratings that can be inserted in the optical path (HETG and LETG, the High- and Low-Energy Transmission Gratings \citep{2005Canizares,2000Brinkman}).  The two focal plane instruments are the Advanced CCD Imaging Spectrometer (ACIS) and the High-Resolution Camera (HRC). ACIS consists of 10 CCD chips and provides images as well as spectral information of the objects/fields observed, operating in the energy range of \mbox{0.2--10 keV} \citep{2003Garmire}. The HRC has two micro-channel plate components and images in the \mbox{0.2--10 keV} range. It also has a time resolution of 16 micro-s \citep{2000Murray}. The ACIS is used either to take high-resolution images with moderate spectral resolution or as a readout device for the transmission gratings.  The ACIS-S3 has a moderate spectral resolution of E/$\Delta$E$\sim$10--30. The High-Energy Transmission Grating Spectrometer (HETGS) works over 0.4--10 keV and has a spectral resolution  ($\lambda$/$\Delta\lambda$) of 60--1000. The Low-Energy Transmission Grating Spectrometer (LETGS) has a range of 0.09--3 keV and a resolution of 40--2000 ($\lambda$/$\Delta\lambda$) .

The \cha\ X-ray telescope is similar to the \xmm \ spacecraft of ESA, launched in the same year, but the two telescopes have different design foci, as \cha\ has a much higher angular resolution, while \xmm\  offers a higher spectroscopy throughput below 2.5 keV (for sources that are not bright enough to be observed with the \cha\ LETG grating). Also, all \xmm\ instruments operate/observe simultaneously. The FOV of \xmm\ is 30$^{\prime}$$\times$ 30$^{\prime}$;
the FOV of \cha\ is variable, and the HRC instrument has a similar FOV, but the ACIS is smaller. The HRC time resolution of \cha\  is about two times better than the lowest EPIC time resolution, and the gratings onboard \cha\ have better resolution.

While the future of \cha \ is at the moment somewhat uncertain, because a portion of funding was withdrawn by NASA in the past year and ``ideas'' have been floating of possibly  having to decommission the HRC camera and the LETG with it, it is important to point out that \cha \ will be unique for many years to come, because (a) even if it has a small effective area, the spatial resolution  is extremely important because it  matches that of UV, IR, and radio instrumentation and ground-based optical telescopes; and (b) the two gratings cover wavelength ranges that cannot be studied with the RGS or with {  the new {\it XRISM}\footnote{\url{https://heasarc.gsfc.nasa.gov/docs/xrism/}}  mission. In particular, the longer X-ray wavelengths of the LETG will not be measurable even in the future with the conceptualized and planned {\it NewAthena}\footnote{\url{https://www.cosmos.esa.int/web/athena}} mission.} 

\section{Cataclysmic Variables}\label{sec:cv}

Cataclysmic variables (CVs) and related systems (e.g., AM CVns, symbiotics),  here referred to as accreting white dwarfs (AWDs),  are compact binaries with white dwarf (WD) primaries (or much less often, because of unstable mass transfer, with WD secondaries). In most CVs, accretion is via Roche lobe overflow, and disk accretion occurs from a donor star that is either a late-type main sequence star or sometimes a slightly evolved star \citep{1995Warner}. CVs have orbital periods of 1.4--15 hrs, with a few exceptions of up to 3 days. Another class of AWDs are the symbiotics, where the donors are red giants or AGB stars.  Accretion in such systems is mainly driven by powerful winds, and most systems show disk formation. Symbiotics have orbital periods ranging from  several hundreds of days to several years. There are exceptions that may be classified as both CVs and symbiotics, in which case the orbital periods are of several days, e.g., \cite[][]{2020Munari}. {AM CVns, another related system to CVs, are ultracompact systems with binary periods between 5 and 65 min that have passed the CV period minimum.
The minimum orbital period in the CV distribution is 65 min as calculated theoretically \citep{1999Kolb,2001Howell}; however, observations indicate a minimum period of \mbox{79.6 $\pm$ 0.2 min} \citep{2019McAllister}. 
AM CVn stars also display Roche lobe overflow with the possibility of stream impact accretion leading to hot spots on the WDs. AM CVns and symbiotic systems will be elaborated upon in Section~\ref{sec:amcvn} and Section~\ref{sec:wdsim}, respectively}.

CVs can be further divided into two main categories \citep{1995Warner}. If an accretion disk forms and reaches all the way to the WD surface, as in the case where the magnetic field of the WD is weak or negligibly small ( B $<$ 0.01 MG), such systems are referred to as nonmagnetic CVs. {They comprise the majority of CVs, and are characterized by their eruptive behavior such as dwarf novae (DNe; CVs with disk outbursts), nova-likes (NLs; high-accretion-state CVs), and classical/recurrent novae (CNe/RNe; explosive burning of the WD envelope); see \citep{2017Mukai,2020Balman,2022PageShaw}}. The other category includes magnetic CVs (MCVs), divided into two subclasses depending on the degree of synchronization of the binary. Polars have strong magnetic fields in the range of 20--230 MG, which cause the accretion flow to be directly channeled onto the magnetic pole/s of the WD, inhibiting the formation of an accretion disk. The magnetic and tidal torques cause the WD rotation to synchronize with the binary orbit. Polars show strong orbital variability at all wavelengths and show multi-wavelength variability on different timescales~\citep{1998Schwope}. Intermediate polars (IPs), which are believed to possess weaker magnetic fields of $\sim$1--20 MG, are asynchronous systems; see \citep{2017Mukai,2022PageShaw}. IPs may accrete via a disk or without a disk (disk-less), or in a hybrid mode in the form of disk overflow, which may be diagnosed by spin, orbital, and sideband periodicities at different wavelengths~\citep{1995Hellier,1997Norton,2020deMartino}. The accretion flow in MCVs is channeled along the magnetic field lines in the close vicinity of the WD (i.e., magnetic poles), reaching supersonic velocities and producing a stand-off shock above the WD surface \citep{1973Aizu}. {The post-shock region is hot ($\rm kT$ = 10--50 keV) and cools via thermal bremsstrahlung (i.e., thermal plasma radiation  giving rise to hard X-rays) and line emission,}  and cyclotron radiation emerging in the optical and/or nIR bands. Both emissions are partially thermalized by the WD surface and re-emitted in the supersoft X-rays and/or EUV/UV domains with a blackbody $\rm kT \sim$20--60 eV due to reprocessing~\citep{2002Schwope,2020deMartino}. Some soft X-ray-emitting  IPs also exist at high accretion rates, reaching a relatively hotter range of blackbodies \citep{2007Evans}.

In nonmagnetic CVs, the transferred material forms an accretion disk that reaches the WD surface. Standard accretion disk theory predicts that half of the accretion luminosity emerges from the disk and the other half from the so-called boundary layer (BL) at the interface with the WD surface \citep{1973Shakura,1974Lynden-Bell}. The theoretically expected BL emission is such that during low-mass accretion states, it is optically thin, emitting in the hard X-rays {
 \mbox{ (T$\sim$10$^{7.9}$--10$^{8.5}$ K;~\citep{1993Narayan,1999Popham})}, and for high-accretion-rate states \mbox{($\rm \dot M_{acc}\ge$ 10$^{-9}$--10$^{-9.5}$ M$_{\odot}$ yr$^{-1}$)}, it is optically thick, emitting in the soft X-rays and EUV band \mbox{(T$\sim$10$^{5}$--10$^{5.6}$ K; \citep{1995Popham,1995Godon,2015Hertfelder})}.}  
It has been shown that optically thin BLs can be radially extended, advecting part of the energy to the WD as a result of their inability to cool \citep{1993Narayan}. The standard disk is often found to be inadequate for modeling high-state CVs,  namely nova-likes (NLs; accreting at a few $\times$10$^{-8}$~\msun yr$^{-1}$ to a few $\times$10$^{-9}$ \msun\ yr$^{-1}$) in the UV as well as some eclipsing quiescent and erupting dwarf novae. In fact, it generates a spectrum that is bluer than the observed spectra, indicating that a hot optically thick inner flow of the BL is found missing \citep{2007Puebla,2010Linnell}. Recent standard disk models have used a truncated inner disk, e.g.,~\citep{2017Godon}, to model the UV data adequately. Standard disk models cannot accommodate and explain the extent and/or emission characteristics of the X-ray region (e.g., hard X-rays) in nonmagnetic CVs. {The accretion flows in high-state nonmagnetic CVs (and DNe) have been explained in the context of} radiatively inefficient (advective) hot flows associated with the hard X-ray emission from these systems as opposed to standard optically thick accretion flows that form in the inner disk, which explain most of the complexities in the X-rays and other wavelengths \citep{2012Balman,2020Balman,2022Balman}. Shock formation in ADAF flows (advection-dominated accretion flows) around AWDs has been calculated \citep{2021Datta}; shocks can occur at a distance of \mbox{$\sim$1.3 $\times$ 10$^9$ cm} from the WD surface, explaining the dominant hard X-ray emission in nonmagnetic CVs at any accretion rate.

\subsection{The Achievements of \cha  }\label{cha:cv}

The first studies of MCVs and nonmagnetic CVs mainly involved \cha\ spectroscopy with the HETG or ACIS (no gratings), which allow the investigation of emission models {via spectral analysis and identification of emission lines. As a result, calculating temperatures and densities via line diagnosis, testing high-resolution data against different plasma models (also under different ionization conditions) can be performed where proper maximum X-ray temperatures, accretion rates, fluxes, luminosities, and  abundances can be obtained in a plasma with a temperature distribution, in turn revealing a proper accretion geometry. The next two subsections will deal with \cha\ studies of accreting nonmagnetic CVs and MCVs. Other types of AWDs and novae are covered in separate sections further in this~review.} 

\subsubsection{Nonmagnetic CVs}\label{cha:nmcv}

Given the moderate- to high-resolution spectra of the ACIS instrument and the additional HETG grating for bright sources,  some DNe were studied  both in quiescence and in  outburst, e.g., SS Cyg and U Gem \citep{2002Szkody,2006Guver}, where the unprecedented spectral resolution revealed  prominent narrow emission lines of O, Ne, Mg, Si, S, and Fe. The line fluxes, ratios, and widths indicated that {quiescent X-ray emission} in U~Gem arose from a range of temperatures in a high-density ($>$10$^{14}$ cm$^{-3}$) gas, moving at low ($<$300 km s$^{-1}$) velocity, with a small ($<$10$^7$ cm) scale height compared to the WD radius. Simple models with cooling flows, cooling flows plus isothermal zones, and thermal conduction showed reasonable agreement with the low-temperature emission lines, but not for a global fitting of the HETG spectrum. \cha\ observed the DN WX Hyi with the HETG which detected strong evidence of outflows/wind from a quiescent DN  for the first time with a velocity of a few $\times$10$^{3}$ km s$^{-1}$ \citep{2003Perna}; however, no other DN has been found to show such a structure during quiescence as of the writing of this review.  The X-ray spectrum displayed strong and narrow emission lines of N, O, Mg, Ne, Si, S, and Fe.  WX Hyi differs from other dwarf novae observed at minimum in having much stronger low-temperature lines, which prove difficult to fit with existing models; the various ionization states implied by the lines suggest that the emission is produced in a wide temperature range, from 10$^6$ to 10$^8$ K. Line diagnostics indicate that most of the radiation originates from a very dense region, with n$_e$$\sim$10$^{13-14}$ cm$^{-3}$. 

\begin{figure}[H]
\includegraphics[height=4.3in,width=5.3in,angle=0]{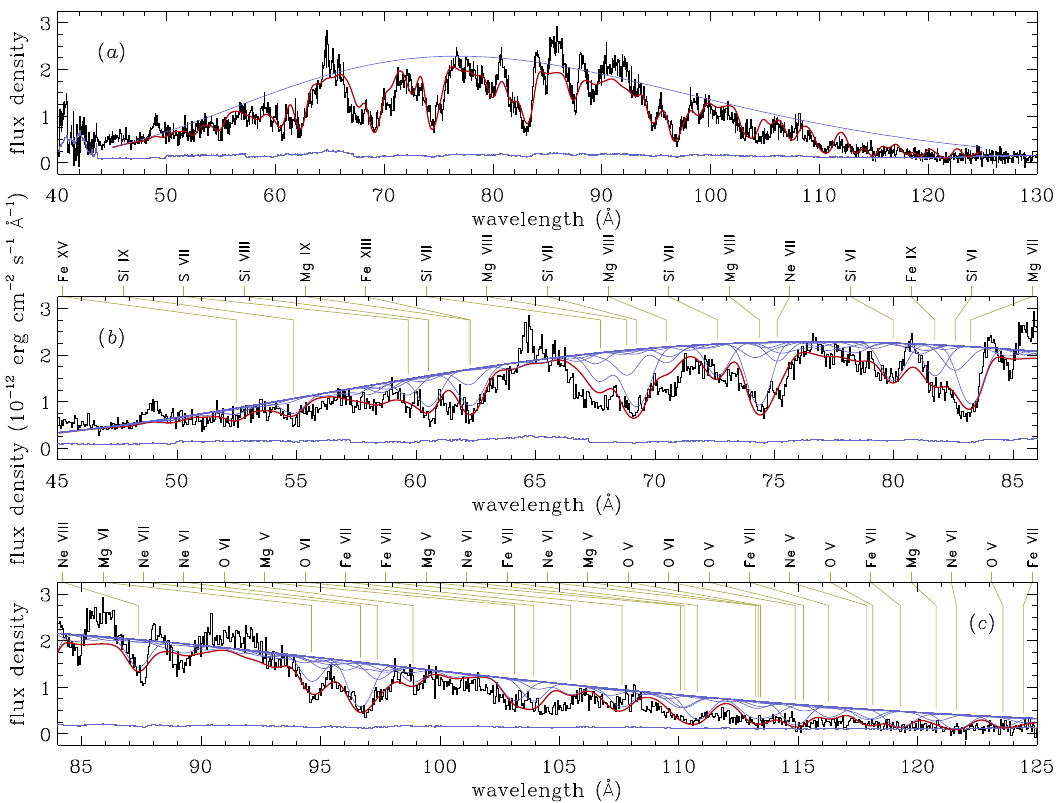}
\caption{\cha\ LETG spectrum of SS Cyg in outburst; figure obtained from \citep{2004Mauche}, reproduced with permission by AAS. It is dominant blackbody emission along with superimposed absorption features (lines, edges). \label{fig:ss-letg}
}
\end{figure}

Using data obtained in June and July, 1994, with the Extreme Ultraviolet Explorer\footnote{\url{https://heasarc.gsfc.nasa.gov/docs/euve/euve.html}}, (EUVE) deep survey photometer and in January 2001 with the \cha\ X-Ray Observatory LETG, \citet[][]{2002Mauche-qpo} investigated the extreme-ultraviolet (EUV) and soft X-ray oscillations of the dwarf nova SS Cyg in outburst, finding QPOs with frequencies of $\nu_0$ = 0.012~Hz and $\nu_0$ = 0.13~Hz in the EUV and $\nu_0$ = 0.0090~Hz and $\nu_0$ = 0.11--0.12~Hz in the soft X-rays using LETG. These findings have extended the Psaltis, Belloni, and van der Klis\ $\nu_{high}-\nu_{low}$ correlation \citep{1999Psaltis} for neutron stars and black hole Low-Mass X-ray Binaries (LMXBs) nearly two orders of magnitude in frequency with $\nu_{low}$$\simeq$0.08$\nu_{high}$.{ This relation suggests that similar mechanisms produce QPOs in compact objects and that dwarf nova oscillations (DNOs) are the counterparts of kHz QPOs, and the low-frequency QPOs are equivalent to the horizontal branch oscillations of LMXBs.} The same data yielded the first high-resolution soft X-ray spectrum  (LETG) of DN in outburst \citep{2004Mauche}. The LETG spectrum (see Figure~\ref{fig:ss-letg}) was found to be consistent with a single-temperature blackbody of T = (2.5 $\pm$ 0.5) $\times$ 10$^5$ K affected by photoelectric absorption and the scattering opacity of an out-flowing ionized wind showing absorption features (20 ions of O, Ne, S, Si, and Mg were recovered with \nh~= (7.9 $-$ 3.5) $\times$ 10$^{19}$ cm$^{-2}$). {This component is believed to be a good example of optically thick boundary layer emission as suggested by theoretical predictions (see Section~\ref{xmm-nmcv}).} The associated boundary layer luminosity was L$\simeq$5 $\times$ 10$^{33}$ \lumcgs\ with a radiative efficiency of L$_x$/L$_{disk}$$\sim$0.05.

Nonmagnetic CVs that have been studied with the HETG, e.g.,
Nova V603~Aql, V426~Oph, SU~UMa, SS~Cyg, and U~Gem, were all observed in quiescence and the last two also in dwarf nova outbursts. Results of a study by the authors of \cite{2006Rana} on the Fe K$\alpha$ emission lines from a sample of such systems indicate the presence of a reflection component with equivalent width consistent with reflection from the WD surface. The Fe~XXV triplet at 6.7 keV is dominant in all sources, with a G-ratio\footnote{G = (f + i)/r; ``f'' is the forbidden, ``i'' is the intercombination and ``r'' stands for the resonance line of the triplet emission where G-ratio reveals the electron temperatures and ionization condition in the plasma.} revealing collisional equilibrium in quiescence with dominant resonance lines. The Fe~XXV line is significantly broadened in U~Gem and SS~Cyg during outbursts, indicating high-velocity material near the WD. The Fe~XXVI/Fe~XXV line ratio points to a higher ionization temperature during outburst. {In outburst, U~Gem shows  the  He-like Fe triplet lines broadened by  2500 km~ s$^{-1}$, and SS~Cyg shows high-velocity material (with a similar calculation), $\sim$1200 km~s$^{-1}$.}

{ In a publication by \cite{2014Schlegel}, the authors summarize the results of a line-by-line fitting} analysis of the available spectra obtained by the HETG data of CVs in general. They confirm the existence of broad ranges of electron temperature and density in all subtypes. The temperatures range from 0.4~keV to 10~keV,  and  single-temperature models do not describe the line emission. The electron density also covers a broad range, from 10$^{12}$ to 10$^{16}$ cm$^{-3}$. The above authors demonstrated that much of the plasma is in a non-equilibrium state (NEI plasma),  but the iron emission may arise from plasma in collisional ionization equilibrium (CIE). Moreover, Comptonization of soft photons by hot electrons present in a cloud surrounding the central source in a transition layer (TL) has been suggested for the X-ray spectra of nonmagnetic CVs in the 0.4--150 keV range \citep{2020Maio}. \xmm\ EPIC pn, \cha\ HETG/ACIS and LETG/HRC, and \rxte\ PCA (Proportional Counter Array) and HEXTE\footnote{ \url{https://heasarc.gsfc.nasa.gov/docs/xte/learning\_center/hexte.html}} observations have shown consistency with one- or two-temperature Comptonized plasma emission models, with a theoretically expected X-ray photon index of 1.8 for the nonthermal emission. A similar model was adopted by the same authors for the IP spectra, yielding a photon index range of 1.3--2.0.  

\begin{figure}[H]

\includegraphics[height=4.0in,width=5.3in,angle=0]{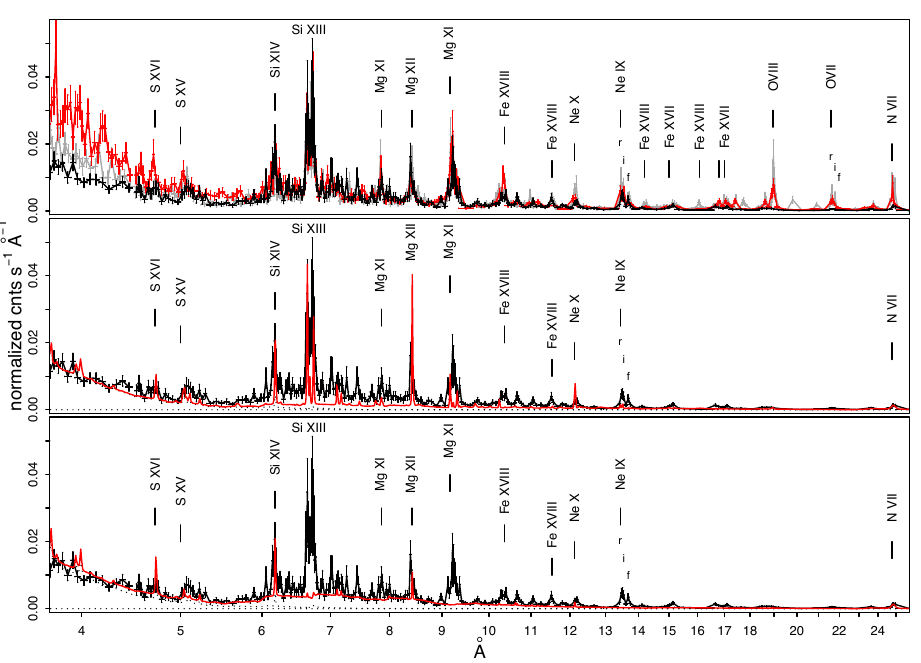}
\caption{ \cha\ HETG spectrum of GK Per obtained during the outburst in 2015 \citep[figure taken from][]{2017Zemko}. {In the top panel, black lines represent 2015, and the gray and red lines indicate the 2002 outburst spectra. The middle and bottom panels show fitted HETG spectra (2015 data) with two different plasma temperatures using a CIE model emission (middle: 0.9 keV, bottom: 4.9 keV). All detected emission lines are labeled.} \label{fig:gk-hetg}
}

\end{figure}

\textls[-15]{Though few, some known MCVs also exhibit DN outbursts (for details, see \mbox{Section~\ref{xmm-nmcv}}).} The DN outburst of GK~Per (IP) in 2015 was observed within a multi-wavelength campaign including \cha\ HETG, where prominent emission lines, especially those from Ne, Mg, and Si, were detected. {A joint \nustar\ and HETG spectral fitting analysis revealed \mbox{16 $\pm$ 0.5 keV} for the post-shock temperature in the accretion column during outburst, which is about only (60--65)\% of the quiescent X-ray temperature. This indicated that increasing ram pressure, as a result of higher accretion rate during the outburst (at the magnetospheric boundary), pushed the accretion disk towards the WD surface, causing a decrease in the shock temperature as the accretion column likely became shorter and denser.} The line diagnosis with HETG ratios of H-like to He-like transition for each element showed a much lower temperature than the underlying continuum, suggesting that the X-ray emission in the 0.8--2 keV range originates from the magnetospheric boundary \citep{2017Zemko}. The results are shown in Figure~\ref{fig:gk-hetg}, where the 2015 and 2002 outburst HETG spectra are compared: the VAPEC\footnote{\url{https://heasarc.gsfc.nasa.gov/xanadu/xspec/manual/node134.html}}, model  (a collisional equilibrium plasma emission model with variable abundances) underestimates the intercombination and forbidden lines in all the triplets (see the middle and bottom panels). Thus, the overall spectrum below 2 keV cannot be represented with a model of plasma in CIE {such that neither one- or two-temperature VAPEC nor a cooling-flow-type plasma can be fitted well, hinting at the existence of NEI conditions and/or a hybrid plasma (exhibiting both photoionization and collisional ionization).}

\subsubsection{Magnetic CVs}\label{cha:mcv}

\begin{figure}[H]

\includegraphics[height=4.0in,width=5.3in,angle=0]{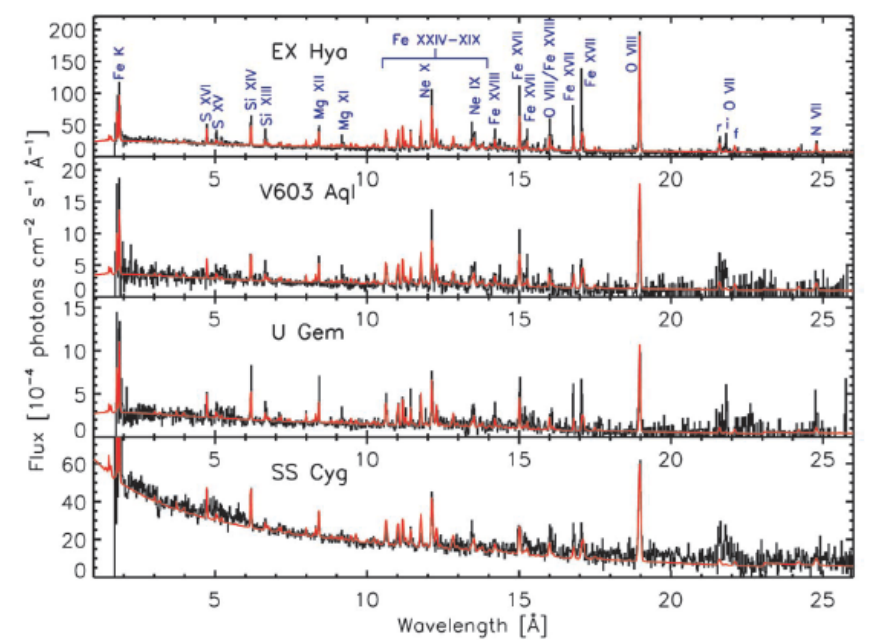}
\caption{ \cha\ HETG spectra of selected CVs fitted with the cooling-flow-type temperature distribution model. Figure obtained from \citep{2003Mukai}, reproduced with permission by AAS. All detected emission lines and the particular source names are labeled.  \label{fig:hetg-sp}
}

\end{figure}

Most studies of MCVs (and nonmagnetic CVs as in the previous subsection) involve \cha\ spectroscopy with HETG looking for correct emission models, line identifications, and spectral parameters of the X-ray-emitting plasma, revealing proper accretion geometry. An early study by \citet{2003Mukai} tested the high-resolution data of CVs against simple isobaric cooling flow (i.e., multi-temperature plasma) and photoionized plasma models. They found that such models show consistency with different sources where maximum plasma temperatures\footnote{($dEM=(T/T_{max})^{\alpha -1} dT/T_{max}$); $\alpha$ = 1 is standard cooling flow, $EM$ is the emission measure \citep{1996Arnaud}.}, accretion rates, luminosity, and abundances can be measured in a plasma with temperature distribution and/or photoionization effects using, e.g., He-like triplets (Figure~\ref{fig:hetg-sp} shows identified lines and fitted spectra). In another following study, the iron K$\alpha$ region of the spectra in MCVs is studied with the HETG \citep{2004Hellier}, which allowed a high resolution for the H-like, He-like, and fluorescent components, with evidence of structure within them. For example, the different shapes of the He-like components in AM~Her (polar) were attributed to their greater cyclotron emission compared with AO Psc. Doppler shifts in the H- and He-like components of MCVs were not detected, indicating that the X-ray emission is predominantly from the denser, lower-velocity base of the accretion columns. 
{A detailed HETG analysis of the prototype polar AM~Her showed H-like and He-like lines of Fe, S, Si, Mg, Ne, and O, with several Fe L-shell emission lines \citep{2007Girish}. The forbidden lines in the spectrum were generally weak, whereas the H-like lines were stronger, as would be expected from a CIE plasma. Phase-resolved spectroscopy showed that the line centers of Mg XII, S XVI, the resonance line of Fe XXV, and Fe XXVI emission were modulated by a few hundred to 1000 km s$^{-1}$ from the theoretically expected values, indicating the bulk motion of ionized matter in a single-pole-accreting binary system.}
\citet{2012Beuermann} improved the soft X-ray spectral model for polars  based on the analysis of the LETG spectrum of AM~Her which involves an exponential distribution of the emitting area vs. blackbody temperature. The model yields more precise values of the wavelength-integrated flux of the soft X-ray component,  and of the implied accretion rate, than reported previously.

The ``Rosetta Stone'' of IPs, the brightest of all, EX~Hya, has been observed with the \cha\ gratings to study the detailed physical parameters that describe the shock zone (e.g., T, n$_e$) and the system parameters via line diagnosis, in the X-rays for the first time. The Fe XXII I(11.92\AA)/I(11.77\AA) density diagnostic applied to the \cha\ HETG spectrum of EX~Hya revealed that this plasma is formed in a region with an electron density of \mbox{(3.0--0.5) $\times$ 10$^{14}$ cm$^{-3}$} and a $T_e$ of about 12 MK \citep{2003Mauche}, which is significantly cooler than the shock region. The radial velocity curve of a cataclysmic variable was measured in the X-rays for the first time by combining line profiles from different portions of the observation \cite{2004Hooger}, deriving a velocity amplitude of 58.2 $\pm$ 3.7 km~s$^{-1}$. 
A follow-up observation of this source with the HETG during 500 ks (divided in four segments) revealed narrow and broad components in the line  profiles with velocities of about 500 km s$^{-1}${  and } 1600 km s$^{-1}$, respectively. The broad component is attributed to the pre-shock region of the accretion flow, photoionized by the post-shock radiation (e.g., O VIII, Fe XVII, Mg XII, Si XIV). By modeling the line profiles, ref. \citep{2010Luna} constrained the height of the standing shock in the range 0.019--0.62~R$_{\rm WD}$. This is re-estimated using \nustar\ data along with \cha\ to less than 0.9R$_{\rm WD}$, and the reflection amplitude\footnote{This is the reflection scaling factor parameter of the {\it reflect} model in XSPEC. It is a convolution model for reflection from neutral material according to the method of \citep{1995Mag}. The X-ray emission from magnetic CVs can change as a result of the reflection effect from the WD surface or accretion curtain which can cause a Compton Hump in the range of 10-30 keV, and the fluorescent Fe K$\alpha$ emission line at 6.4 keV \citep{2018Hayashi}. For example, a tall accretion column reduces the solid angle of the WD, reducing also the Compton hump at large viewing angles, which can explain the low reflection amplitudes less than 1.} is found less than 0.15. This long data set is also used to update the mass estimate to 0.79 \msun\ and to validate the standard cooling flow model for the post-shock accretion column radiation. The results indicate that even if modifications to the standard model {are introduced,} variable gravitational potential along the column, magnetic pole geometry, departures from constant pressure, resonant scattering, different electron-ion temperatures, or Compton cooling cannot explain the observed line ratios in this source, where perhaps thermal conduction modifies the temperature distribution \citep{2015Luna}. The complexity of the magnetic accretion flow and the shock zone is thoroughly revealed and studied in detail only with \cha.

The enigmatic source AE~Aqr (IP), which is a magnetic propeller (the only known propeller system for decades until 2022), was observed with the HETG (and also \xmm\ RGS) to understand the X-ray accretion geometry and characteristics \citep{2009Mauche}.   The accretion column model appeared to be unsupported,  as it failed to reproduce the Chandra HETG X-ray light curves and radial velocities, and it required an X-ray luminosity that was more than two orders of magnitude greater than observed in the 0.5--10 keV,  to heat the UV hot spots by reprocessing. The flux ratios of He-like triplets in both the \xmm\ RGS and in the \cha\  HETG spectrum  indicated that either  the electron density  increased with temperature by over three orders of magnitude, from 6 $\times$ 10$^{10}$ cm$^{-3}$  to \mbox{$\simeq$1 $\times$ 10$^{14}$ cm$^{-3}$}, or the plasma was significantly affected by photoionization. The \xmm\ \ RGS \ results showed H-like K$\alpha$ lines broadening by  1250--1600 km s$^{-1}$  and a hot plasma with $kT_{max}$$\sim$$4.6$ keV \citep{2006Itoh}. The radial velocity  measured in the HETG spectrum  emission lines was modulated with  the white dwarf spin phase, with two oscillations per spin cycle and an amplitude of K$\sim$160 km s$^{-1}$. These results are inconsistent with the prior models of AE~Aqr, which were thought to be an  extended, low-density source of X-rays; they support instead earlier models in which the dominant source of X-rays was the high-density one and/or was in close proximity to the white dwarf.

In a different study, a Compton-downshifted shoulder of the fluorescent line of GK~Per was recovered \citep{2008McNamara}. A study of XY~Ari (IP) using two long \cha\ ACIS-S CCD observations at moderate spectral resolution recovered/confirmed orbit- and spin-phased {sinusoidal behavior in the covering absorption column, and found that the iron line components (6--7 keV) require different ionization states (parameters) with log($\xi$) $<$ 2 (6.4 keV line) and log($\xi$) = 3.5--4.0  (6.7 and 6.9 keV lines) that vary particularly with the spin phase}~\citep{2004Salinas}. 

Recently, for the first time, gravitational redshift  has been revealed, using the \cha\ HETG,  in  an MCV, RX J1712.6-2414 \cite{2023Hayashi}. {It was  measured  for the  H-like Mg, Si, and S lines as $\Delta$E/E$_{rest}$$\sim$7 $\times$ 10$^{-4}$ (Mg, Si), and 15 $\times$ 10$^{-4}$ (S), using a weak gravitational regime with $c\Delta$E/E$_{obs/rest}$ = 0.635 \msun/\rsun. This also provides a new method to estimate the WD mass, in this case resulting in M$_{\rm WD}$ $>$ 0.9 \msun\ (cf. Figure 3 in \citep{2023Hayashi}).}

\subsubsection{Detection of Old Novae and Their Shells}

{Classical novae (CNe) outbursts occur in AWDs as a result of an explosive burning of accreted material on the WD surface (Thermonuclear Runaway---TNR) causing the ejection of 10$^{-7}$ to 10$^{-3}$ \msun\ of material at velocities up to several thousand kilometers per second~\citep{2008Bode}. The classical nova systems have an initial low-level accretion phase ($\le$10$^{-10}$~\msun\ yr$^{-1}$), where the recurrent novae (RNe) that are found in outburst with 20-100 years of re-occurrence time generally show higher accretion rates at the level of 10$^{-8}$--10$^{-7}$\msun\ yr$^{-1}$ \citep{2013Anupama}. Studies on novae  with \cha\ and \xmm\ will be described in Section~\ref{sec:nova}.}

\begin{figure}[H]
\begin{adjustwidth}{-\extralength}{-4.4cm}
\centering
\includegraphics[height=2.in,width=5.7in,angle=0]{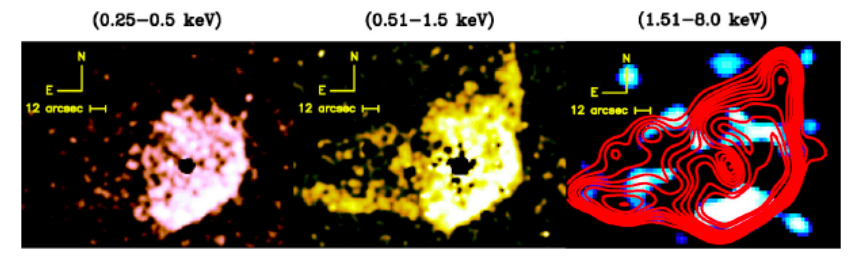}
\includegraphics[height=2.in,width=4.0in,angle=0]{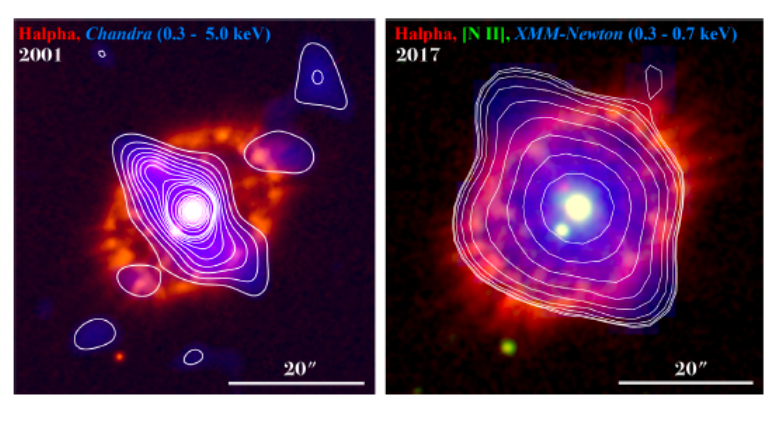}
\end{adjustwidth}
\caption{{ Upper panel shows the nova shell of GK~Per detected with the \cha\ ACIS-S ($\sim$100 yrs after eruption) in different energy bands. Figure is obtained from \citep{2005Balman}}. The central source is removed, and the brightness contours of the entire X-ray image are overlaid on the highest-energy image for comparison. The bottom-left-hand panel shows  the extended emission in the form of elongated X-ray jets associated with the old nova shell of DQ Her. The contours show the X-ray flux  measured with ACIS-S  overlaid on the optical image in H$\alpha$, obtained with William Herschel Telescope in 2001. The bottom-right-hand panel shows  the Nordic Telescope (NOT)  H$\alpha$  image (2020) and  the \xmm\ EPIC pn  contours of the extended emission,  larger and different from  the \cha\ nebulosity. { Both of the bottom panels are obtained from \citep{2020Toala}}. \label{fig:nova}
}

\end{figure}

Some  old novae systems have been observed with the ACIS-S CCDs to investigate their spectral and  temporal characteristics long after eruption; e.g., see also Section~\ref{sec:nova}~\citep{2005Mukai}. {Particularly, spatially extended emission has been investigated with the expectation that the expelled fast ejecta would interact with the surrounding environment. The unprecedented spatial resolution of \cha\ ACIS for imaging analysis can be exploited to recover extended emission from relatively nearby (or relatively older) novae.} The full remnant (\mbox{$\sim$1--1.5 arc min}) of the nova shell of GK~Per was observed and resolved by  \citet{2005Balman} about 100 yrs after explosion. This study revealed nonthermal emission from the shell, with a photon index of 2.3 and a luminosity of 4.5 $\times$ 10$^{30}$ \lumcgs, along with two thermal components (kT = 0.1--0.3 and kT = 0.5--2.6). The nonthermal emission is due to diffusive nonlinear particle acceleration at the SW, where the remnant interacts with a molecular dust cloud and is decelerated. The remnant shows a wedge-like shape with a faster flow in the NW to SE direction and expansion velocities $\sim$2600 km s$^{-1}$. The top panels in Figure~\ref{fig:nova} show the morphology in different energy bands. A distinct emission line of neon, He-like Ne IX, is detected, revealing several emission clumps/blobs. The X-ray luminosity of the forward shock is 4.3 $\times$ 10$^{32}$ \lumcgs. The shocked mass, the X-ray luminosity, and comparisons with other wavelengths suggest that the remnant started cooling and is most likely in a Sedov--Taylor phase. {A  second paper on  the expansion and evolution of the remnant of GK~Per \citep{2015Takei} finds that the flux decline is evident in fainter regions and the mean decline is 30--40\% in the 0.5--1.2 keV energy band with a typical expansion of the brightest part of the remnant as 0.$^{\prime\prime}$14 yr$^{-1}$,
so the fading of the X-rays is due largely to expansion (i.e., the plasma temperatures did not significantly change since the 2000 epoch).}

After \citep{luna2009} accidentally revealed the presence of an anomaly in the \cha\ PSF, \citep{juda2010,montez2022} used sub-pixel-level deconvolution techniques to reveal extended soft X-ray emission detected even 5 years after the 2006 outburst of the recurrent nova RS~Oph. The emission was asymmetric, oriented in the East--West direction of the sky. The sequence of \cha\  observations allowed the measurement of an expansion velocity of 4600 km s$^{-1}$ (D/2.4 kpc) in the plane of the sky, consistent with the optical and radio observations. In a nutshell, the X-ray-emitting plasma is not cooling, but expanding freely in the polar directions into a cavity left by the 1985 eruption.

The recurrent nova T Pyx was observed prior to the last outburst in 2010 (in quiescence), and the central source was detected with kT $>$ 37 keV at about 1.0 $\times$ 10$^{32}$\lumcgs. This is explained in the context of advective (ADAF-like) hot flows \citep{2014Balman-aa}\ since the system is known to have a high accretion rate in quiescence with an expected high accretion luminosity \mbox{(several $\times$ 10$^{35}$ \lumcgs).} Extended emission at S/N of 7--10 has been recovered using the same observation utilizing deconvolution at the sub-pixel level, revealing an elliptical/ring-like shape with an outer radius of $\sim$0.9 arcsec reminiscent of likely the 1966 outburst \citep{2014Balman-aa}.
  
  { In a publication by \cite{2020Toala}, the authors find }extended emission associated with the nova remnant of DQ~Her in the form of a bipolar jet-like structure extending 32 arcsec NE to SW direction using \cha\ ACIS-S (see Figure~\ref{fig:nova}, bottom-left-hand panel). The \xmm\ observation also revealed additional presence of a diffuse X-ray emission from a hot bubble filling the nova shell (see Figure~\ref{fig:nova}, bottom-right-hand panel). The bipolar feature can be modeled by an optically thin plasma emission at low temperature and a power-law component with a photon index of 1.1 $\pm$ 0.9 and a luminosity  of (2.1 $\pm$ 1.3) $\times$ 10$^{32}$ \lumcgs.

\subsubsection{Resolved CV Populations in Globular Clusters and Surveys}

High-spatial-resolution ACIS imaging of globular clusters (GCs) along with spectral analysis revealed several new CVs and related objects. Globular clusters host a variety of lower-luminosity (L$_x$ $<$ 10$^{35}$\lumcgs) X-ray sources. A comprehensive catalog of more than 1100 X-ray sources in 38 Galactic globular clusters observed by the \cha\ ACIS detector identified a substantial underabundance of bright (L$_x$ $>$ 10$^{33}$\lumcgs) IPs in globular clusters,  compared to the Galactic field, in contrast with the literature of the past two decades \citep{2020Bahramian} {hinting at formation-channel problems}. A study on 47~Tuc (GC) indicates an apparent lack of short-period CVs below the period gap. This may be attributed to a high occupation fraction of nonmagnetic CVs with an overabundance of long-period CVs hosting a subgiant donor, and a substantial fraction of CVs within the period gap, with a steep radial surface density profile. These are best understood if many of the CVs have been recently formed via dynamical interactions in the dense cluster core \citep{2023Bao}, {which contrasts with CVs being old populations.}

The ChaMPlane\footnote{\url{https://www.cfa.harvard.edu/research/champ-chandra-multiwavelength-project-and-champlane-chandra-multiwavelength-plane-survey}} project is a multi-wavelength survey designed to classify the serendipitous X-ray sources  detected within the Galactic plane. This survey lead to well-defined spatial parameters (space density and scale height) for the CV distribution within a well-determined volume \citep{2005Grindlay,2009vandenberg,2012Servillat}. Assuming a CV distribution with a scale height of 160 pc, the ChaMPlane observational results are best fitted assuming a local space density of CVs of (0.4--2.4)$\times$10$^{-5}$ pc$^{-3}$ at the 95\% confidence level \citep{2008Rogel}.

{The galactic ridge coincides with the galactic plane of the Milky Way, which shows apparently diffuse X-ray emission concentrated in the galactic plane, named galactic ridge X-ray emission (GRXE). After decades of attempting to resolve the GRXE, long \cha\ observations ($\sim$12 d) revealed that 80\% of the emissions (473 sources) were resolved to be AWDs (ave. 0.6 \msun) and stars with active coronae \citep{2009Revnivtsev-nat}. 
However, more recent works suggest that GRXE has a diffuse component (80 keV) arising from the scattering of the radiation of bright X-ray binaries (XBs) by the interstellar medium \citep{2010Turler}. An \xmm\ Slew Survey result finds that the GRXE is consistent with both diffuse and point-source emissions \citep{2014Warwick}. \suz\footnote{\url{https://www.isas.jaxa.jp/en/missions/spacecraft/past/suzaku.html}} observations yield spectra successfully fitted by the sum of two components using MCVs of mass of 0.66 $\pm$ 0.09 \msun\ and a softer optically thin thermal emission of a plasma with kT in the range of 1.2--1.5 keV that can be attributed to coronal X-ray sources \citep{2012Yuasa}. \citet{2022Schmitt} find that such sources are optically faint and more consistent with AWDs: MCVs and DNe (utilizing {\it Gaia}).} 

\subsection{Fundamental \xmm\ Contributions}

\subsubsection{Nonmagnetic CVs}\label{xmm-nmcv}

The Disk Instability Model DIM \citep[][]{2001Lasota,2020Hameury} predicts that DN outbursts\footnote{Ongoing accretion at a low rate (quiescence) is interrupted every few weeks to months or sometimes longer durations by
intense accretion (outburst) of days to weeks where $\dot{\rm M}$ increases. Super-outbursts are observed in some DNe which are rarer and occur with larger magnitude differences between the quiescence and outburst.} are caused by thermal--viscous instabilities resulting in the brightening of accretion disks that traces the state change in CVs with outflows primarily in the optical and UV during the outburst \citep{2020Balman,2024Tampo}. The DNe in quiescence and outburst have been observed with both \xmm\ (more pointings/studies achieved) and  \cha\ (discussed in Section~\ref{cha:nmcv}). 

During the outburst stage, since accretion rates are higher, at 10$^{-10}$--10$^{-8}$ M$_{\odot}$ yr$^{-1}$ \citep[][]{2011Knigge}, 
the BL is expected to be optically thick, emitting EUV/soft X-rays as blackbody radiation; see the X-ray reviews in \citep[][]{2006Kuulkers,2017Mukai,2020Balman}. 
Such soft X-ray/EUV components with temperatures in a range between 5 and 30 eV are detected only from about five systems compared to all DNe in CVs,
e.g., \citep[][]{2000Mauche,1996Long,2004Mauche,2009Byckling,2023Kimura}, where none of the detections\footnote{We note that the blackbody emission in the DN-like outburst of symbiotic system T CrB was detected with \xmm.} belong to \xmm.
The total X-ray luminosity during outburst is generally in the range of 10$^{30}$--several $\times$ 10$^{34}$ erg s$^{-1}$, where the maximum limit is observed when the soft component is detected (note that \mbox{\citet{2023Kimura}} measured 10$^{35}$ erg s$^{-1}$). On the other hand, as a {persistently} recovered emission component (all DNe), DNe show hard X-ray emission, detected with \xmm\ and other observatories, during the outburst stage at diminished flux levels (X-ray suppression during the optical peak phase) and at lower X-ray temperatures compared with the quiescence level, e.g., \citep[][]{2003Wheatley,2004McGowan,2009Ishida,2010Collins,2015Balman,2017Zhang,2023Dutta,2023Dobrotka,2024Balman}. Only a few DNe show increased levels of hard X-ray emission in outburst, detected with \swi\ and \cha, e.g., GW~Lib and U~Gem \citep[][]{2009Byckling,2006Guver,2021Takeo}{, which may be linked to increased accretion rate.}

\begin{figure}[H]
\includegraphics[height=4.5in,width=5.3in,angle=0]{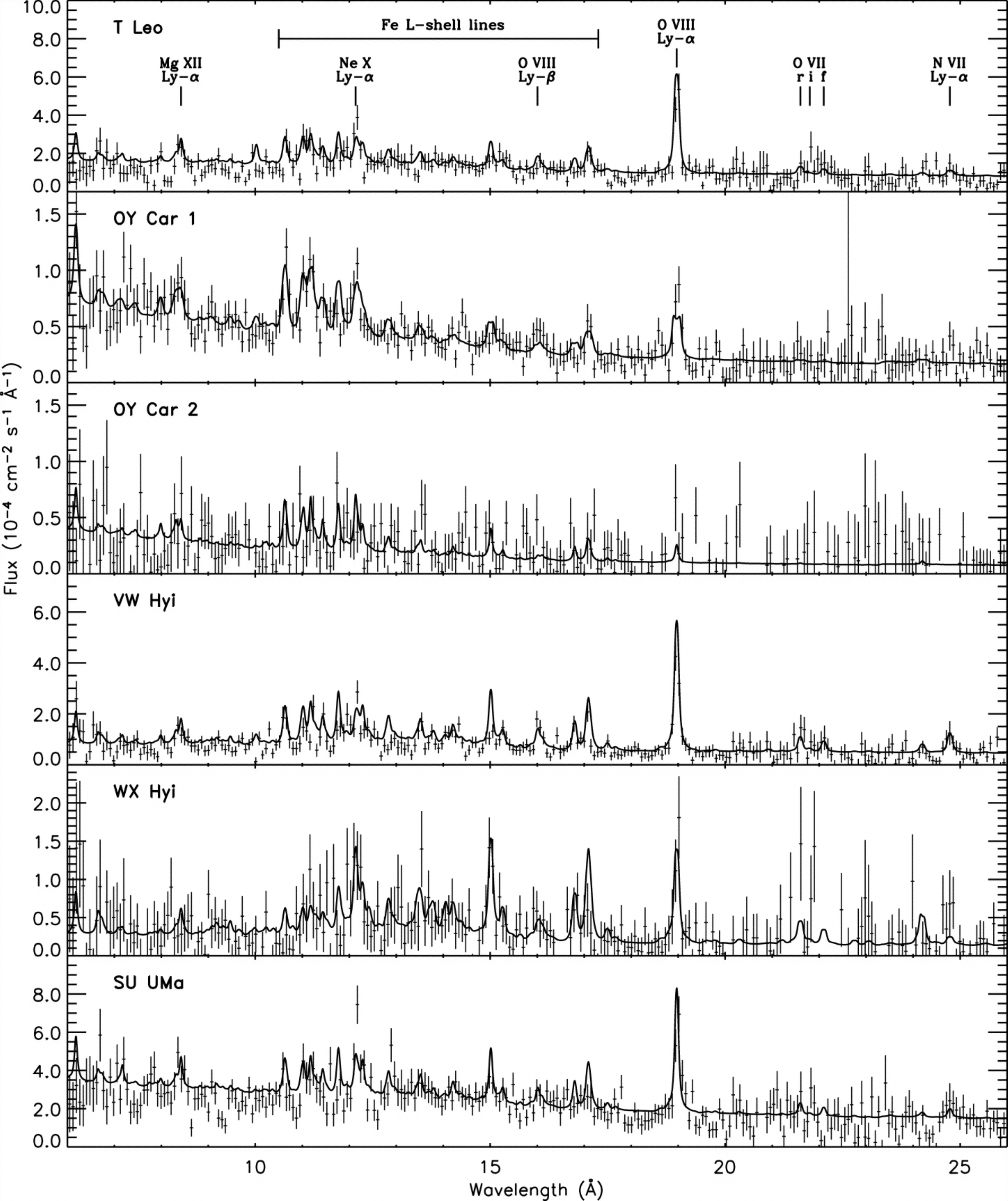}
\caption{ \xmm\ RGS spectra of selected DNe in quiescence with labeled emission lines. Figure is taken from \citep{2005Pandel}, reproduced with permission by AAS. Note that the lines are weak and narrow. The O VIII emission line is the strongest. \label{fig:pandel}
}
\end{figure}

DNe in low-$\dot{\rm M}$ states were observed with \xmm\ and several other X-ray telescopes, including \rxte, \cha, \suz, \swi, \nustar, and \nicer, and the spectral results generally  seem in accordance with optically thin BLs in standard steady-state disks in quiescence at 8--55 keV plasma temperatures. {A multi-temperature cooling-flow-type plasma emission spectrum (well-fitted to EPIC Spectra) is characteristic with mostly H- and He-like emission lines of C, N, O, Ne, Mg, Si, and Fe, in addition to the Fe L-shell lines (O VIII line is the strongest line recovered) in the high-resolution RGS spectra of \xmm\  observations, e.g., \citep[]{2001Ramsay,2003Pandel,2004Hakala,2005Pandel,2007Hilton,2009Nucita,2011Balman,2014Nucita,2024Balman}. Figure~\ref{fig:pandel} shows the quiescent RGS spectra. We note that the lines are weak and narrow, limited by the RSG spectral resolution (line widths \mbox{$<$1000 km s$^{-1}$)}.} 

\begin{figure}[H]
\begin{adjustwidth}{-\extralength}{-4.4cm}
\centering
\includegraphics[height=5.6in,width=5.0in,angle=0]{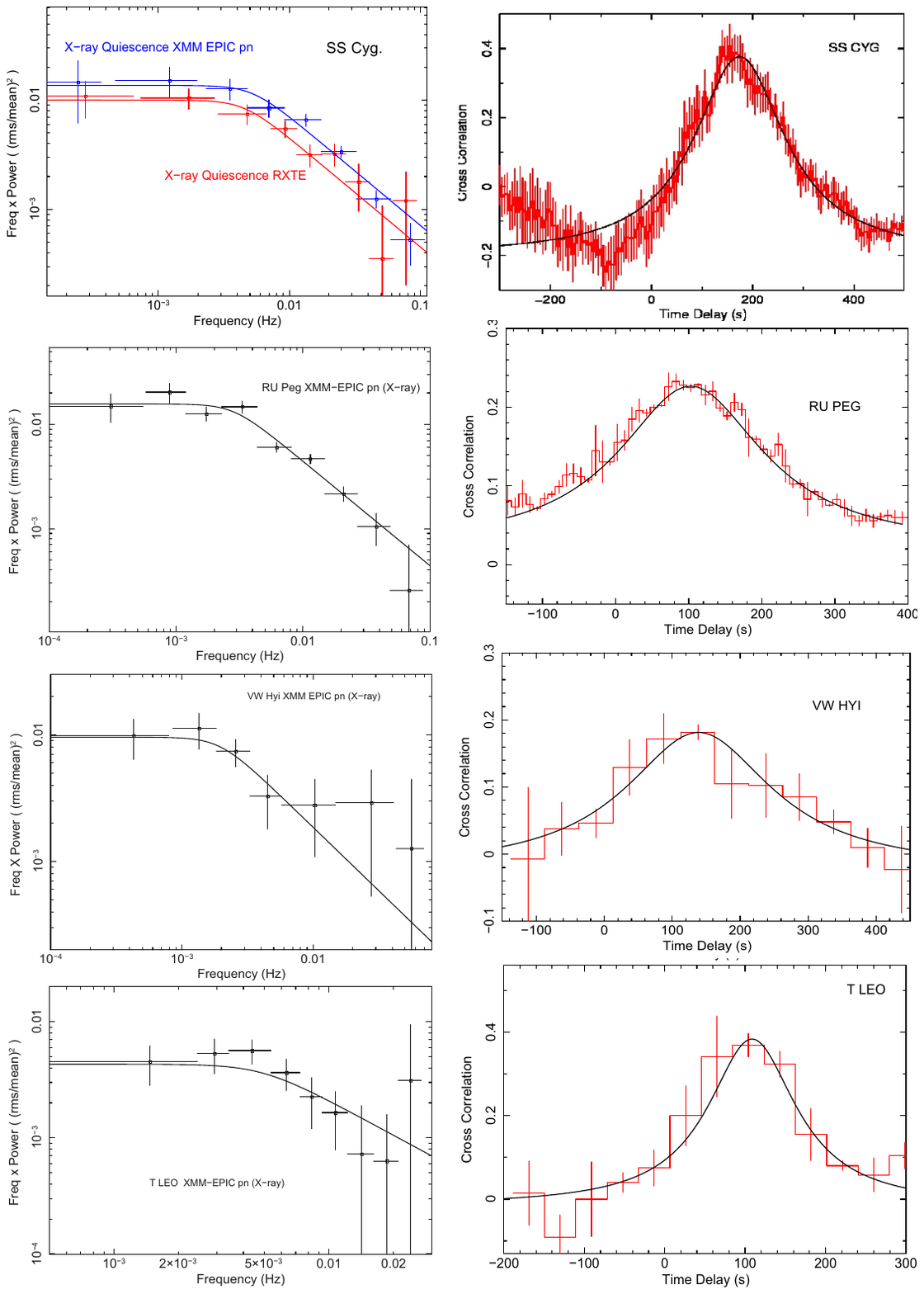}
\end{adjustwidth}
\caption{\xmm\ power density spectra (PDS) of selected DNe in quiescence that show the frequency breaks on the left-hand column, and on the right-hand column is the X-ray and UV cross-correlation of \xmm\  time series data. 
The 90--180 s X-ray lags are evident. See the original paper for  figures and details of the analysis \citep{2012Balman}, reproduced with permission by ESO. \label{fig:balman}
}
\end{figure}

On the other hand, studies of X-ray variability via power spectral analysis of noise using mainly \xmm\ time series data (and some \rxte) have shown that the inner disk structure of DNe in quiescence reveals optically thick disk truncation and formation of hot coronal flows in the inner parts of the quiescent DNe accretion disks; these structures do not disappear during outbursts, revealing hard X-ray components \citep{2012Balman,2019Balman,2020Balman,2023Dobrotka}. 
In these studies of aperiodic time variability, the broadband noise power spectra are derived. The break frequencies detected in the characteristic power-law red noise structure show the change in and the diminishing Keplerian flow of a standard Keplerian disk into a sub-Keplerian advective hot flow. The range of break frequencies is 1--6 mHz for quiescent dwarf novae, derived from \xmm\ data alone, translating to a transition radius of (3--10) $\times$ 10$^{9}$ cm \citep{2012Balman,2020Balman,2023Dobrotka}. \xmm\ OM (UV or B-band) power spectra indicate a similar break frequency range in quiescence. Detailed cross-correlation analyses using \xmm\ time series data of quiescent DNe show 90-180 s lags of X-rays with respect to UV (\xmm\ OM fast time series data), indicating propagation delays and change in flow structure \citep{2012Balman,2020Balman} into advective hot flows within DN disks. {As a result, radiative inefficiency is also revealed in the X-ray luminosities, only relatively in quiescence, but more prominently in outburst. Figure~\ref{fig:balman} shows examples of quiescent DNe PDS with the frequency breaks (on the left) and examples of cross-correlation calculations indicating the lags on the right.}

\begin{figure}[H]
\includegraphics[height=2.45in,width=5.45in,angle=0]{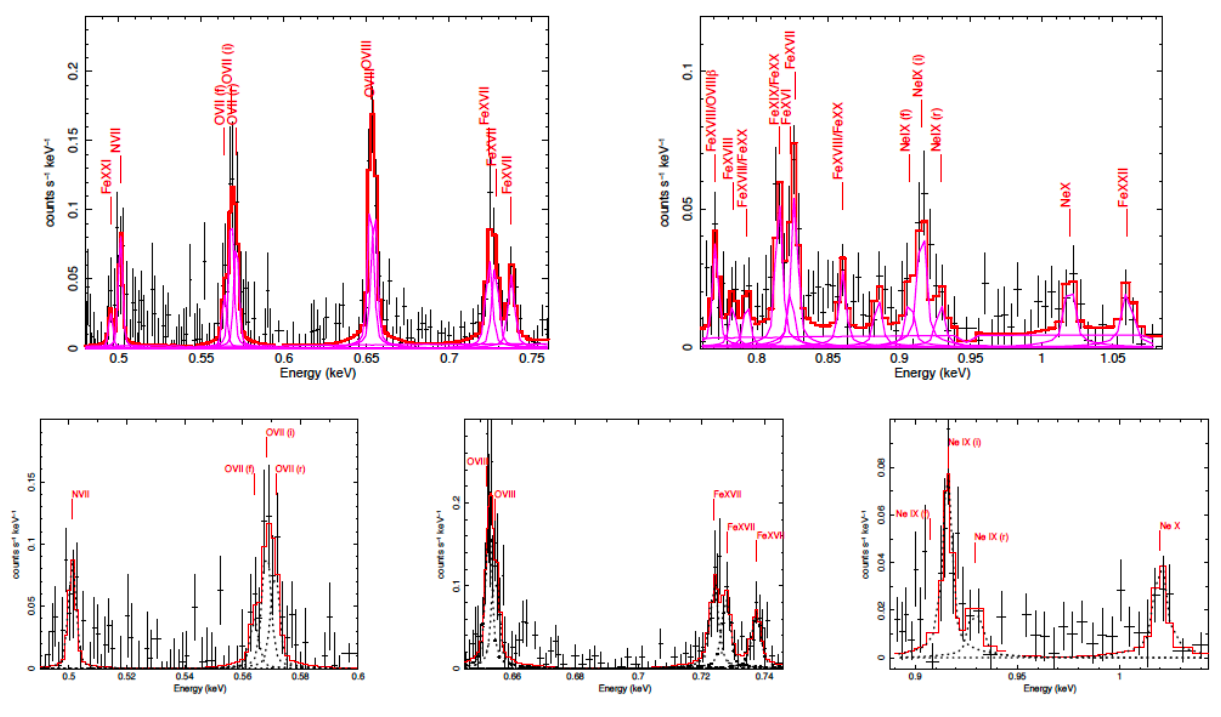}
\includegraphics[height=1.47in,width=5.45in,angle=0]{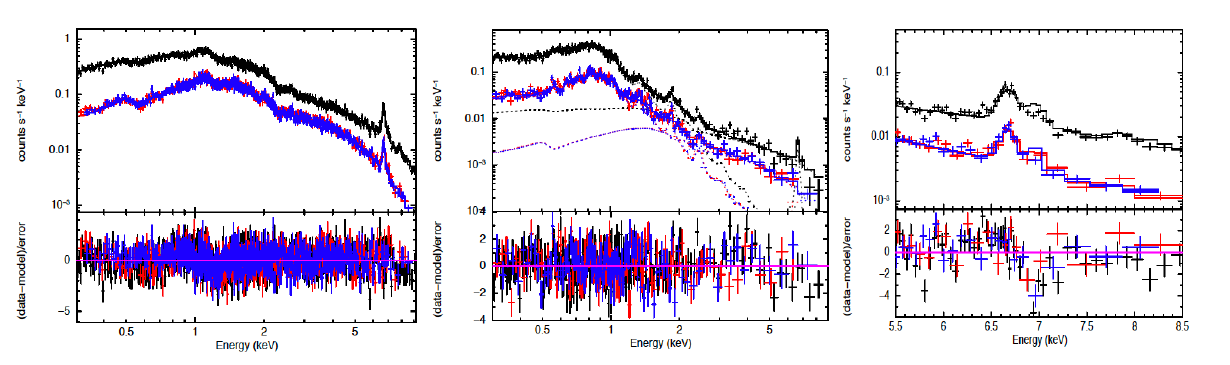}
\caption{{ Upper panel and the middle panel show the \xmm\ RGS line diagnosis of DN Z Cha in outburst \citep{2024Balman}. All detected and fitted emission lines are labeled.} The bottom panel displays the fitted EPIC spectra of Z Cha, in quiescence and in outburst (on the left and in the middle of panel). The right-hand side (bottom) is a close-up on a fitted quiescence EPIC spectra that show the Fe XXVI absorption line. { All figures (in these panels) are obtained from  \citep{2024Balman}, reproduced under CC-BY 4.0 license AAS.} \label{fig:sb_spec}
}
\end{figure}

Thus, time series analysis of \xmm\ data has introduced a new understanding of the disk structure and of the nature of the accretion flows in DNe (also in NLs---see the next paragraph and perhaps in other AWDs). In addition to this, a recent study by  \mbox{\citet{2024Balman}} analyzed the \xmm\ EPIC  and RGS spectra of  the DN Z Cha with  detailed line diagnostics  in quiescence and outburst,  showing that the X-ray-emitting region is extended (vertically and horizontally) in both  source states. There is some or no collisionally ionized plasma  in equilibrium. The  low radiative efficiency, particularly in the outburst ($\sim$0.0004; L$_x$/L$_{disk}$), with the  electron densities of (7--90) $\times$ 10$^{11}$ cm$^{-3}$, indicate advective hot flows both in quiescence and outburst in the inner disk of the DN instead of a basic BL picture (see Figure~\ref{fig:sb_spec} for RGS line identifications, fits, and the fitted EPIC spectra). This study also detected the first Fe XXVI absorption line in an accreting CV in quiescence (Figure~\ref{fig:sb_spec}, bottom-right-hand panel), with partially ionized absorbers in quiescence {(ionized absorber has an equivalent \nh~= (3.4--5.9) $\times$ 10$^{22}$ cm$^{-2}$ and a log($\xi$) = 3.5--3.7)} and outburst. This shows that warm absorbers are/can be typical components of nonmagnetic CVs in general. {Ionized absorbers were used to fit nonmagnetic CV spectra obtained with some past 
missions, e.g., \asca~ \citep[][]{2005Baskill}, and consistent results were found with an ionization parameter log($\xi$) in the range of $-$1.5--2.1.  \xmm\ with a higher sensitivity of the EPIC pn and relatively better spectral resolution than \asca\ can 
yield good results in assessing ionized absorbers in nonmagnetic CVs (as well as MCVs; see Section~\ref{xmm:mcv}) as in the case of Z Cha.}

As mentioned above, the \xmm\ RGS spectra of most quiescent DNe reveal only narrow emission lines,  with line broadening $<$1000 km s$^{-1}$, inconsistent with the rotational broadening expected from BLs \citep{2003Pandel,2005Pandel,2024Balman}, but consistent instead with sub-Keplerian advective hot flows as suggested \citep{2020Balman,2024Balman}. Moreover, the long-term \xmm\ data of VW Hyi have shown a clear decline trend in time {of  flux since the last outburst} \citep{2019Nakaniwa}.  
As similarly found in the \rxte\ analyses of SS~Cyg \citep{2004McGowan} and SU~UMa \citep{2010Collins}, the declining fluxes in between outbursts contradict the predictions of the expected flux increase in the DIM model. It is consistent with truncation of the optically thick flow structure in the disks as the outbursts subside into quiescence (as the hot flow returns to its quiescent form). The duty cycles of DNe have been found to vary   between 30\% and 0.6\% in the X-rays using the 12 \xmm\ observations and a calculated  X-ray luminosity function in quiescence \citep{2010Byckling,2015Britt}. One DN is found to have a high-duty cycle of 50\% (AX J15498-5416;~\citep{2017Zhang}). { We caution that due to the recovered advective nature of the X-ray emission in general of nonmagnetic CVs, such calculations (e.g., duty cycles, luminosity functions) may not be~reliable}. 

Some high-state CVs, NLs, have been observed with \xmm\ and a few also with \cha\, \swi\ and \nustar. These observations have not detected  the expected optically thick soft X-ray component (i.e., blackbody emission) just like in the older generation of X-ray telescopes, but only hard X-ray emissions were detected \citep{2014Balman-apj} along with absorption in the orbital plane for some systems (orbital variations even at low (orbital) inclination angles were found) \citep{2004Pratt,2010Hoard,2014Page}. The optically thin hard X-ray emission with virial temperatures and non-ionization equilibrium conditions in the X-ray-emitting region in NLs, plus the lack of soft X-ray emission at high states, together with similar broadband noise characteristics to DNe, has been interpreted as existence of ADAF-like radiatively inefficient advective accretion flows near the WD; \citep[see][for reviews]{2020Balman,2022Balman}. 
{This indicates that NLs resemble other X-ray binaries in disk flow structure and emission (though with some differences).} From an analysis of the Kepler data and \xmm\ data of the NL MV~Lyr, it was also proposed that in the very inner region, the geometrically thin, optically thick disk is surrounded by a geometrically thick and optically thin disk region, a “sandwich” model which may explain the hard X-ray emission and the flickering variability (though it would suffer from large-scale Comptonization effects) \citep[][]{2017Dobrotka}. 

\subsubsection{Supersoft X-ray Sources}

The supersoft X-ray sources (SSSs) were discovered with the \Einstein\ Observatory and have been established as a new class of X-ray binaries on the basis of observations with \rosat. They have extremely soft spectra with blackbody emission at temperatures of 15--100 eV. 
They are highly luminous with bolometric luminosities of 10$^{36}$--10$^{38}$ \lumcgs. Their observed characteristics are consistent with WDs, which are steadily or cyclically (with on/off phases) burning hydrogen-rich matter accreted onto the surface at a rate of around 10$^{-7}$ \msun\ yr$^{-1}$. The required high accretion rates can be supplied by mass transfer on a thermal timescale of 10$^{6}$--10$^{7}$ yrs from close companion stars that are more massive than the white dwarf accretor around 1.3--2.5 \msun; see \citep[][]{1997Kahabka,2006Kahabka-b,2013Orio} for a review. Steady H-burning can also occur in a nova outburst stage, but for shorter timescales, observed in classical novae and symbiotic novae (see Section~\ref{sec:nova}). SSS systems are important in evolutionary scenarios { for stable burning at the highest accretion rate} regimes and for unstable mass transfer phenomena, and finally, they are expected to be Supernova (SN) Type-1a progenitor candidates.

SSSs have been observed during the \cha\ and \xmm\ eras. The main improvements have come from the detailed \cha\ LETG spectroscopy, along with the \xmm\ RGS, which have allowed us to study system spectral characteristics, geometry, etc. The prototypical SSS source CAL 83 in the Large Magellanic Cloud (LMC) revealed a very rich absorption-line spectrum from the hot white dwarf photosphere with no spectral signatures of a wind \citep{2005Lanz}. In this study, the LETG and RGS spectra of CAL 83 were matched with  new NLTE (non-local thermodynamic equilibrium) line-blanketed model atmospheres, which used LMC abundances, yielding the first direct spectroscopic evidence that the WD in this system is massive with 1.3 $\pm$ 0.3 \msun\ along an effective temperature of 550,000 $\pm$ 25,000 K. Another { well-known} LMC SSS, CAL 87, has been observed by LETG and RGS at differing epochs, which yield similar spectral results, with complex and unusual line ratios. A recent investigation finds that continuum X-ray flux is at least an order of magnitude too low for a hot H-burning WD, but it is consistent with Thomson-scattering reflecting 5\% of the WD emission spectrum, with an effective temperature of 800,000 K and a mass of $\sim$1.2 \msun~\citep{2024Pei}. A large Thomson-scattering corona explains the X-ray eclipse of CAL 87, in which the size of the eclipsed region is on the order of a solar radius. The emission lines originate in an even more extended region beyond the eclipsed central X-ray source. Overall, several results support that the wind-driven mass transfer scenario is running in CAL 87 \citep{2024Pei,2014Ribeiro}. { These results indicate that the two well-studied SSS sources CAL 83 and CAL 87 share very different disk-flow and wind structure.}

Magellanic Cloud (MC) surveys with \xmm\ on finding new SSS have yielded interesting sources within the class: the EPIC-pn spectrum of XMMU J050803.1--684017 has been fitted using a blackbody spectrum with an effective temperature of (26--51) eV at a bolometric luminosity of (0.1--30) $\times$ 10$^{36}$ \lumcgs, depending on the assumed absorption. The source is consistent with the nucleus of a planetary nebula \citep{2008Kahabka}. { A handful of SSS in the Magellanic Clouds (sample size is increasing in time)} that could not be identified with transient nova outbursts were found to be massive close binaries (strong H$\alpha$, H$\beta$, and He II emission lines) consisting of Be stars and WDs \citep{2006Kahabka,2025Marino,2012Li}  by optical identification e.g., \citet{2018Cracco}. An evolutionary scenario can be found in \citet{2023Gies}. If these systems are abundant, this may be a new channel for the explosion of type Ia SNe 
in star-forming regions, without invoking double-degenerate systems as progenitors. These binary systems { have been predicted} decades ago by \citet{2001Ragu} to constitute 70\% of all Be binaries. WDs are expected to be embedded in the Be-star disk envelope; thus, it may be difficult to detect the X-rays because of absorption; however, this can explain SSS being driven at a high accretion rate onto the WD, resulting in high luminosities and possibly unstable mass transfer. 
Note that the very soft spectra and the high luminosities in excess of 10$^{34}$ \lumcgs\ cannot be explained in the framework of accretion in nonmagnetic CVs (they are not H-burning and are at lower-accretion-rate regimes; see Section~\ref{xmm-nmcv}). 
In general, \xmm\ surveys have detected several SSSs in galaxies M31, M33, LMC, and Small Magellanic Cloud (SMC) (aided by their low extinction), where most of these are identified as novae \citep{2006Misanovic,2011Stiele,2014Henze}.

\subsubsection{Magnetic CVs}\label{xmm:mcv}

The first observations of MCVs with \xmm\ of IP or polar-type were made by pointed observations of well-known systems that had yet to be explored for spectral and temporal characteristics (e.g., plasma and blackbody temperatures, luminosities, power spectra, and periodicities) that were not fully derived by previous missions with less sensitivity and/or spectral and timing resolutions \citep[][]{2001Ramsay,2002Ramsay,2002Schwope-dp,2004Ramsay,2004Evans,2005Rana,2005Schwarz,2006Evans,2007Norton}. These studies have also been able to demonstrate detailed geometrical modeling of the accretion column, the post-shock zone (PSR), the polar hot spots, the accretion disk, and the accretion impact region. Moreover, some of these studies involve low-state polars \citep{2002Pandel,2004Ramsay-low,2005Pandel-low,2007Schwope,2009Schwarz} since \xmm\ provides a good sensitivity for such states. In general, polars change their accretion state to a low state with a 3--4 magnitude difference in their V magnitude {(intermediate states with \mbox{1.5--2 mag} difference exist).} The luminosities in the low state are found to be $\sim$a factor of 100 lower than the high state utilizing \xmm~\citep{2004Ramsay-low}. Surveys of polars using \xmm\ and \swi\ show that the soft X-ray components are absent in low states with detected hard X-ray plasma emission from the stand-off shock with temperatures $\le$5 keV at luminosities $\le$$1\times10^{30}$~\lumcgs\ (note that a recent \nicer\ study \cite{2024Balman-aa} {finds a blackbody component in the low state (15--18 eV) along with a harder X-ray component at higher X-ray temperatures $\sim$8.5 keV).}  In addition, some pre-polars have been recovered and investigated with \xmm\ (using also UV and optical spectroscopy and photometry). These systems are the progenitors of polars, with the secondary not yet in Roche lobe contact and the WD accretes from the stellar wind of the secondary with $\dot{\rm M}$ $<$ 10$^{-13}$ \msun yr$^{-1}$. The X-rays show plasma emission at $\le$10$^{29}$ \lumcgs with kT $\le$ 1 keV  \citep[][and references therein]{2004Szkody,2011Vogel}. 
The origin of the X-rays has been suggested to be from coronal emission on the secondary star. 

A puzzle during the older-generation missions' time-frame, particularly in the \rosat\ period, was that the soft-to-hard X-ray luminosity ratio revealed a soft X-ray excess in the blackbody-emitting MCVs, namely, polars.  This problem was resolved with the aid of \xmm\ using 21 systems \citep{2004Ramsay-eb}. Reanalysis of the \rosat\ data with newer calibration files  reduced  the magnitude of the soft excess, which  may now be explained as a  result of the reflection of hard X-rays from the WD surface, optical thickness corrections, and the cyclotron emission component. However,  the ``blobby accretion'' model  \citep{2000King} that explains the soft excess emission is still viable, since at high accretion rates, blobs are expected to  penetrate the shock and thermalize in the atmosphere, creating a large soft excess (traditional bombardment solution; \citep{1988Frank,2004Ramsay-eb}). This scenario requires at least L$_{soft}$/L$_{hard}$$\sim$10 in the X-rays and high H column densities. 

\begin{figure}[H]
\includegraphics[height=2.47in,width=5.45in,angle=0]{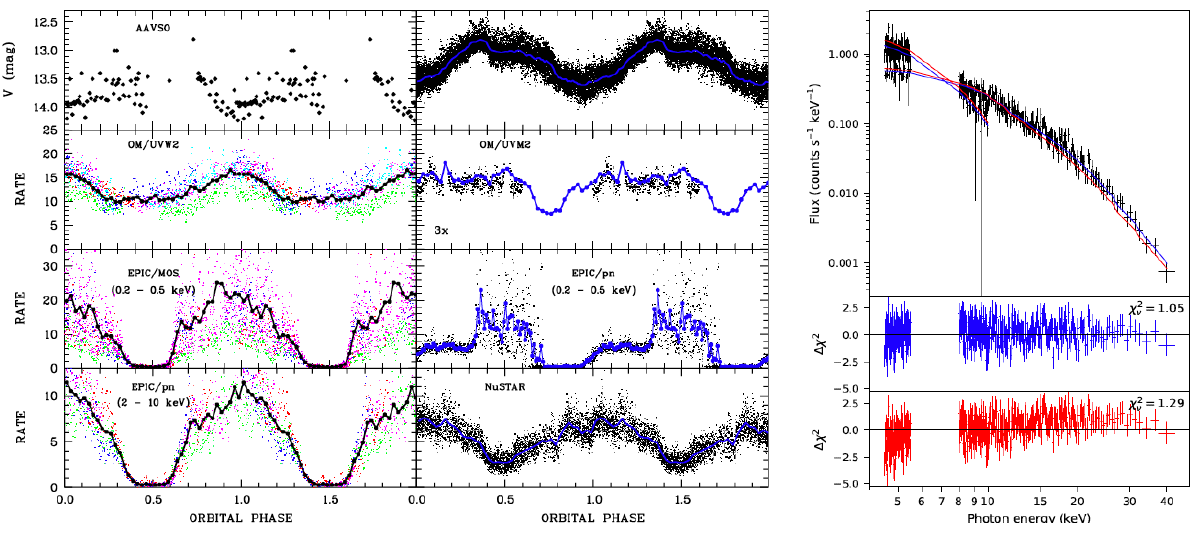}
\caption{{ Left-hand panel shows  the ``regular mode'' light curves of AM Her and the middle panel shows the ``reversed mode'' light curves including \xmm. The figure is obtained from \citep{2020Schwope}. } The right-hand panel  shows the combined \xmm\ and \nustar\ spectra,  fitted using the reflection model {\it reflect} (in XSPEC) along with a plasma emission model. Different colors present fits with or without (red) {\it reflect} model indicating significance of reflection~\citep{2020Schwope}, reproduced with permission by ESO. \label{fig:schwope}
}
\end{figure}

An optically thick soft X-ray component arising from the heated polar cap is known to characterize polars. {The \xmm\ surveys and pointed observations revealed an excessive number of new IPs, instead, showing soft blackbody emission with effective temperatures in the range of 40--100 eV \citep{2020deMartino}, which increased the three \rosat-discovered systems to 19, 8 of which were also found to be polarized \citep{2007Evans,2008Anzolin,2017Bernardini}. These systems have a soft-to-hard X-ray luminosity ratio that is lower than that of polars. In general, \xmm\ sensitivity did not provide an excessive number of new soft X-ray components for polars, but provided similar statistical characteristics to \rosat\ with temperatures in the range of 20--60 eV \citep{2020deMartino}.} On the other hand, \xmm\ has shown a remarkably increasing number of polars without the soft X-ray excess/component, suggesting a group of hard X-ray polars { \citep[][and references therein]{2019Bernardini}.} 
{Polars are also known to exhibit various accretion modes; a detailed multi-wavelength study of the enigmatic polar AM~Her (mostly in high states),  utilizing  broadband spectra of \xmm\ and \nustar\  \citep{2020Schwope}, showed X-ray emission from the main pole that exhibits self-eclipse, revealing the soft X-ray component from the secondary pole in the ``regular mode'' (observed in 2005). The system showed a ``reversed mode'' observed in 2015 (first detection in \citep{1994Paerels}) with strong blobby (soft component) accretion at a far accretion spot along with a non-eclipsing main accretion region indicating migration of the accretion footprint, resulting in the out-of-phase light curves for the two different modes (see Figure~\ref{fig:schwope}).  Moreover, the \nustar\ spectrum combined with \xmm\ revealed a significant existence of a Compton hump as the result of reflection together with the 6.4~keV emission line.}

\begin{figure}[H]
\includegraphics[height=1.56in,width=5.45in,angle=0]{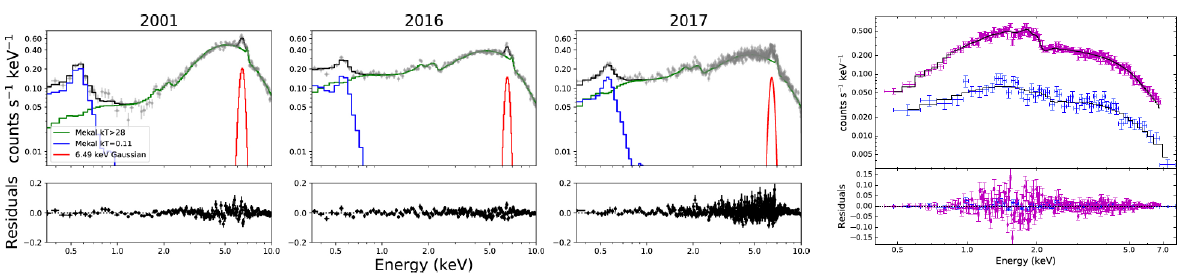}
\includegraphics[height=1.57in,width=5.45in,angle=0]{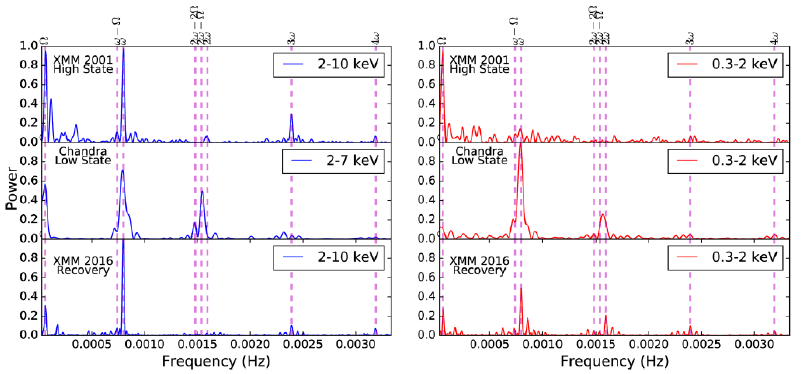}
\caption{Upper panel shows the \xmm\ EPIC pn spectrum  of FO Aqr  in a high/normal state on the left with the recovery states from the low state. On the right-hand side  is the \cha\ low state. The lower panels are the power spectra of these states, as labeled. { All the plots are obtained from  \citep{2017Kennedy,2020Littlefield}, reproduced partly with permission by AAS.} \label{fig:foaqr}
}
\end{figure}

Polars show frequent high and low states, as mentioned before; in contrast, IPs do not. However, a handful of systems (around four) have been caught in optical state changes (by about  at most 2 mag in the V band) \citep{2008Staude,2019Bernardini-st,2020Littlefield}. The best-followed one is FO Aqr, which showed a low state followed by higher states in 2016--2018. Historical low states are also recovered in 1965--1966 and 1974 \citep{2020Littlefield}. FO Aqr is observed with \xmm\ (and with \cha) in 2016--2018. It is suggested that in these low states, the accretion disk is dissipated into a non-Keplerian ring of diamagnetic blobs.  {The observations indicate  that the accretion mode changes from a disk-fed configuration in the high state (spin period-dominated), to a predominantly stream-fed configuration in the low state (with a strong beat period detected, denoted as 2$\omega-\Omega$, in the stream-fed mode, where $\omega$ is the spin period of the WD, and $\Omega$ is the binary period)}.
X-ray spectra show excessive soft X-ray emission (not a blackbody model) following the low states, and the neutral absorption column density drops considerably by 5--10 times during the low state (1.5 $\times$ 10$^{22}$ cm$^{-2}$), where the plasma temperature drops by about two times down to 15~keV (a lower limit by \cha), and the flux drops by a factor of 8--10 times \citep{2017Kennedy} . Figure~\ref{fig:foaqr} shows the low-state spectra along with the recovery spectra and an older high/normal state spectra in the upper panel; the lower panel displays the power spectra of these states {as labeled, where different periods, such as the spin and the beat periods, are dominant depending on the accretion mode.}

Most IPs and polars show energy-dependent pulses, with amplitudes increasing at lower energies, indicative of the photoelectric absorption effect, and the soft excesses reveal partial covering absorbers (perhaps with a distribution) from the pre-shock accretion flow (N$_{\rm H }$$\simeq$10$^{21-23}$ cm$^{-2}$~\citep{2017Mukai}).  Moreover, some systems show a flat distribution or slight increase in pulsed fractions with energy in MCVs. This is expected if electron scattering opacity (e.g., reflection effects) and warm absorber effects take place/dominate in the accretion flow; see \citep[][]{2024Balman-aa,2021Islam} for a detailed explanation. An additional absorption component is required in  those IPs displaying energy-dependent orbital modulations; see \xmm\ analyses by \citep[][]{2011Pekon,2012Pekon,2018Bernardini}, where the accretion impact region seems to be a warm absorber {location with N$_H$$\sim$several $\times$10$^{22}$ cm$^{-2}$ (first indications by \citep{2005Parker}).} For the first time, \xmm\ and \cha\ grating spectra have demonstrated at least in three IPs the presence of an O VII absorption edge \citep{2008deMartino,2012Bernardini}. This indicates that in some IPs (or MCVS, for that matter), the pre-shock flow can also be substantially ionized, and thus, ionized absorber models have been used when appropriate in the modeling of the broadband MCV spectra \citep{2021Islam,2017Mukai}. Moreover, the presence of a Compton reflection hump in MCVs has been found using \nustar\ observations of hard X-ray IPs, e.g., \citep[][]{2015Mukai,2022Dutta}, in combination with \xmm\ spectra for the lower-energy response. This component is not detected in the spectral fits using the low S/N average BAT and/or IBIS-ISGRI spectra of IP systems (along with \xmm EPIC spectra). However, reflection either at the WD surface or in the pre-shock flow is confirmed by the fluorescent Fe at 6.4 keV with equivalent widths (EWs) in the range 100--250 eV using \asca~\citep[][]{1999Ezuka}. 

Many X-ray spectral models of MCVs have been developed that account for temperature and gravity gradients, dominant cyclotron cooling (in polars), solving one- or two-fluid hydrodynamic equations, and dipolar field geometry, e.g.,~\citep[][]{2007Saxton,2014Hayashi}, while modifications for magnetospheric disk truncation, especially in low-field IPs, have been implemented to obtain more reliable mass estimates using the PSR model \citep[][]{2019Suleimanov}. 
Studies focused on calculating the masses of WDs have been among the key studies for MCVs/CVs due to an asserted WD mass problem, where the WDs in CVs presently show high mass averages, such as 0.77 $\pm$ 0.02 \msun, using the PSR model \citep[][]{2020Shaw}, from a \nustar\ study, which is not well accounted for by evolutionary models. 
{We note here that some WD dynamical mass calculations, e.g., for IPs, differ from estimates obtained by modeling the X-ray spectral continuum and measuring X-ray temperatures \citep{2024Alvarez}}. In addition, WD masses have been investigated using flux ratios of Fe XXVI-Ly$\alpha$ to Fe XXV-He$\alpha$ emission lines (I 7.0/I 6.7) from archival \xmm\ and {\it Suzaku} observations \citep{2022Yu}.  They obtain average WD masses of 58 CVs (including 36 IPs and 22 nonmagnetic CVs), yielding similar results for both the magnetic and nonmagnetic species at 0.81 $\pm$ 0.21 \msun . 
\subsubsection{CV/MCV Populations and Surveys}

\xmm\ studies of AWDs  have also made use of  synergy  with optical/IR/UV  surveys/ground-based observatories, most notably the SDSS (Sloan Digital Survey), ZTF (Zwicky Transient Factory), TESS (Transiting Exoplanet Survey Satellite), CRTS (Catalina Real-time Transient Survey), ATLAS, Gaia, and SALT Observatory. Particularly, several  MCVs and nonmagnetic WD systems have been identified as candidates, e.g., in the SDSS, and have been confirmed with \xmm\ pointed observations, e.g., \citep[][]{2005Schmidt,2005Homer,2006Homer,2006Szkody,2009Hilton,2016Worpel}. Some of these systems form part of the handful of period-bounce systems (CVs that have evolved to very low donor masses past the CV period minimum) studied with \xmm\ and recently using the \erosita\ survey, with some confirmed to be MCVs {(they show X-ray orbital modulations and have magnetic WDs)} { \citep[][and references therein]{2023MunozGiraldo}.} 
This work reveals X-ray temperatures of 3--5 keV, X-ray luminosities of 10$^{28}$--10$^{30}$ \lumcgs, and $\dot{\rm M}$ of (3--9) $\times$ 10$^{-14}$ \msun yr$^{-1}$ consistent with the evolved CV nature. 

The statistics of period-bounce systems are important for understanding the evolution of CVs, since these make up the population at the very late stages of evolution  (40--70\% of the CV population) when these systems evolve from the period minimum to longer periods and larger binary separations, as the secondaries resemble brown dwarfs. Only a small  number of such systems are known, but a recent theoretical study explores the reasons for this discrepancy, reconciling it with its predictions \citep{2023Schreiber}.

Magnetic and nonmagnetic AWDs  have been detected as mostly hard X-ray emitters and cataloged in X-ray surveys using \xmm, \swi, and \integral\ (majority being MCVs). They are mostly bright hard X-ray sources with luminosity in the range of 10$^{30}$--10$^{34}$~\lumcgs\ and they play an important role in understanding the Galactic X-ray binary populations for the surveys; see \citep[][]{2022Xu,2020deMartino,2013Heard,2016Hailey,2018Oh}. These studies suggest dominance of MCVs in the X-ray populations, particularly of IP-type above 10$^{31}$ \lumcgs. Though optical spectroscopy and photometric follow-ups have been used to confirm the new MCV candidates, the detection of X-ray pulses and X-ray spectral characteristics are the key diagnosis method for MCVs to confirm existence of magnetic accretion.  Therefore, the role of \xmm\ for MCV studies has been crucial in resolving the magnetic status with  programs (long term) of pointed observations, where around 30 such systems have been confirmed with about 25 IP detections \citep{2012Bernardini,2017Bernardini,2018Bernardini} and about 3 new polar system confirmations { \citep[][and references therein]{2019Bernardini}}. 

\begin{figure}[H]
\includegraphics[height=2.3in,width=5.5in,angle=0]{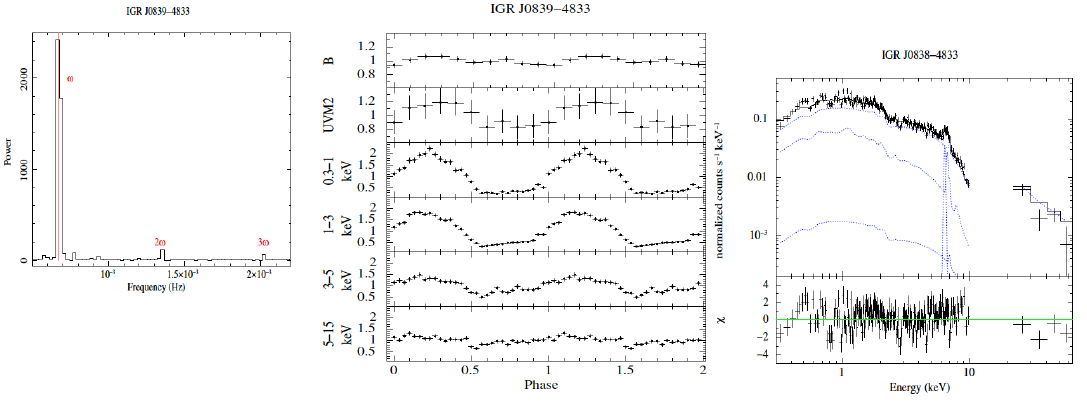}
\caption{ \xmm\ EPIC pn power spectrum of IGR J0839-4833 on the left. The middle panel shows the energy-dependent mean light curves using \xmm\ EPIC and OM (folded at the spin period). On the right is the fitted \xmm\ EPIC pn spectrum combined with INTEGRAL ISGRI data (4 data points above 20 keV) to accommodate the hard/high-energy tail. The lower panel on the right shows the residuals of the fit in standard deviations. Figure panels obtained from \citet{2012Bernardini}, reproduced with permission by ESO. \label{fig:bernardini}
}
\end{figure}

 New IPs show dominant spin periodicity as expected, e.g.,~\citep[][]{2012Bernardini,2009Anzolin}, some with weaker power at the harmonics, indicating the presence of a secondary emitting pole, but purely double-pole accretors (pulses at 2$\omega$), which may also show transient behavior (such as accretion to one or two poles), have been detected and studied with \xmm\ as \mbox{well \citep{2012Bernardini,2022Ok,2022Kennedy}.} Figure~\ref{fig:bernardini} shows an example \xmm\ energy spectrum, power spectrum, and mean light curves for a (new) IP. 
In addition, these IP systems frequently display  the beat  frequency (2$\omega - \Omega$), which indicates the existence of stream overflow on the disk, \mbox{e.g.,~\citep[][]{2012Bernardini,2024Rawat,2024Joshi}}. This is a consequence of the stream impact at the inner Lagrangian point where the stream overflow toward smaller disk radii may occur. The stream is well described by a ballistic trajectory, but larger masses of gas can be swept up and overflow at smaller, but still highly supersonic, velocities \citep{1998Armitage}. As this stream impacts the magnetosphere spinning with the WD, a beat frequency is observed. Disk overflow is also an observed phenomenon studied primarily in the optical wavelengths in nonmagnetic CVs~\citep{2012Smak,1998Schreiber}.

The long uninterrupted \xmm\ observations show that MCVs exhibit substantial orbital variability as well.  \textls[-10]{Previously known IPs were concentrated in a limited range of the spin--orbit period plane, clustering just below P$_\omega$/P$_\Omega$$\sim$0.1, with most systems found above the 2--3 h orbital period gap, typically in the range of 3--6 h. With new identifications, the plane has been substantially populated at long periods with 15 systems at P$_\Omega$ $>$ 6 h \citep{2020deMartino}.}


\section{AM CVn Systems}\label{sec:amcvn}

AM CVn systems are ultracompact binary systems with helium-dominated optical spectra at orbital periods of 5--65 min (past the CV period minimum); see~\citep[][]{2005Nelemans,2010Solheim} for a review. 
These objects have different evolutionary scenarios. One model involves a double-WD system that has reached shorter evolutionary periods as a result of angular momentum loss by gravitational wave radiation. They start mass transfer at orbital periods of a few minutes, and evolve to longer periods with a decreasing mass transfer rate. Another scenario is a low-mass WD and a non-degenerate helium star binary that transfers mass as it evolves to a minimum period of 10 min when the star becomes semi-degenerate. Finally, CVs with evolved secondaries lose their outer hydrogen envelope, and uncover their He-rich core, and then evolve as helium stars \citep{2023Belloni}. {  The observed space density of AM CVn systems is (5.5 $\pm$ 3.7) $\times$ 10$^{-7}$ pc$^{-3}$ \citep{2024Rodriguez,2022Roestel}; however, the theoretical expected densities are about a factor of 10--30 times larger \citep{2001Nelemans,2015Goliasch}. 

Galactic compact binaries with orbital periods shorter than a few hours emit detectable gravitational waves at low frequencies that can be recovered with the LISA\footnote{\url{https://lisa.nasa.gov}} mission. Some AM CVn systems have been identified as  LISA verification binaries \citep{2024Kupfer}, as their GW amplitude and characteristic strain are strong enough to be resolvable by LISA. Moreover, a significant contribution to the LISA Galactic binary background is expected from the Galactic population of CVs \citep{2023Scaringi}.}

AM CVns show time variability and/or distinct transient states/characteristics as a function of increasing orbital period, and as they evolve to longer periods, they go through three distinct phases. One is a high-state phase with \emph{p} $\le$ 20 min (high accretion rates), where the disk is optically thick  
and shows low-amplitude periodicities at the orbital period, the superhump period, and/or their beat period \citep{2002Patterson,2005Patterson}. 
Another one is the quiescent state associated with
low accretion rates and optically thin disk. Such systems/states are associated with systems \emph{p} $\ge$ 40 min and  studied spectroscopically with only some showing
He absorption lines \citep{2005Roelofs} and emission lines (in the optical) visible during the quiescent low-mass-transfer states \citep{2001Groot}. 
There is a common bursting state during  which optical variability is measured with periods between 20 and 40 min. During these phases, the systems resemble the high state and show absorption lines. 

In outburst, these systems are 3--5 mag brighter than in quiescence. Outbursts recur on timescales from 40 d to several years, and  some systems show  super-outbursts, e.g.,~\citep[][]{2011Levitan}.
The outburst characteristics have been modeled with the  DIM model of \mbox{DNe \citep{2012Kotko,2016Hameury,2020Hameury}}, but the changes in outburst patterns for AM CVn systems are not well explained, e.g., CR Boo \citep[][]{2016Isogai}.

Prior to \cha\ and \xmm, AM CVn stars were found to be weak X-ray emitters, where the emission originates from the accretion disk and/or the wind, and they were thought not to host strongly magnetized WDs \citep{1995Ulla,1994vanTeeseling,1996vanTeeseling}.
In addition, they show no coherent or quasi-periodic variability in the X-rays, but they may show  orbital modulations. By means of  \cha\ and \xmm\ observations, further study of detailed spectra, better measurements of the X-ray luminosity, estimate of abundances, improvements on timing noise and periods, and period derivatives calculations have been achieved, e.g.,~\citep[][]{2003Strohmayer,2004Strohmayer-c,2021Strohmayer}, along with constraints on their  binary evolution.  

\cha, \xmm, and \swi\ observations of extremely low-mass WDs with massive companions (about seven systems)  have been studied, and upper limits to their X-ray luminosity/flux \citep{2014Kilic,2016Kilic} have excluded a {neutron star} origin for the primary in such systems. Calculations indicate that these extremely low-mass WD binary systems are progenitors of AM CVns (discovered for the first time with the aid of X-ray observations). In the next 160 MG yrs, these systems are expected to perform stable mass transfer via an accretion disk  and eventually to turn into AM CVns. Following this in a 10$^8$ yr span, they may result in thermonuclear supernovae. 

\begin{figure}[H]
\includegraphics[height=3.6in,width=5.3in,angle=0]{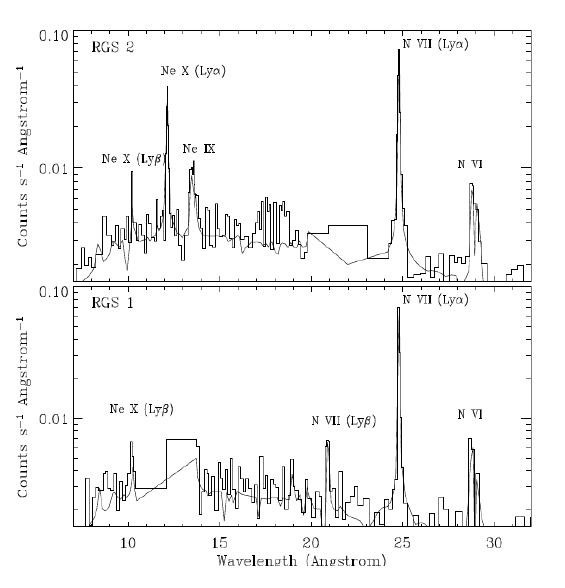}
\caption{ \xmm\ RGS 1 and 2  spectrum of the AM CVn system GP Com. All detected lines are labeled. Figure is taken from \citet{2004Strohmayer-b}, reproduced with permission by AAS. \label{fig:gpcom}
}
\end{figure}

Both \xmm\  and  \cha\ X-ray spectra of AM CVns are modeled with thin multi-temperature thermal plasma models with highly non-solar abundances that
show carbon and oxygen deficiency (CNO-processed material) with significant nitrogen overabundance indicative of nova eruptions  \citep{2005Ramsay,2006Ramsay,2007Ramsay,2004Strohmayer-b,2008Strohmayer}. 
The first X-ray emission lines from a double-degenerate AM CVn (GP Com) were revealed by \xmm\ via the H-like and He-like N and Ne emission lines, where the line ratios revealed  a dense, collision-dominated plasma at a temperature of 6.5 keV with an underabundance of C and O relative to N \citep{2004Strohmayer-b} (see Figure~\ref{fig:gpcom}).
The X-ray plasma temperatures (from about a dozen objects) are found to be cooler than DN in the range of 3.5--8.7 keV. The temperatures in quiescence and off outburst show a power-law distribution in the plasma with ${\alpha}$ = 0.8--1.05 { (power-law index for the temperature distribution of the cooling flow plasma),} typical of DN
in quiescence. No soft X-ray emission from a blackbody is detected from most of  these systems on or off outburst in general, even at high accretion rates, except for the two stream impact systems, V407~Vul (RX~J1914.4 + 2456) and HM~Cnc  (RX~J0806.3 + 1527). The X-ray luminosity of AM CVn systems 
is in the range of 1.5 $\times$ 10$^{32}$--1.7 $\times$ 10$^{30}$ erg s$^{-1}$ but largely found to be L$_x < $ 10$^{32}$ erg s$^{-1}$, which is generally similar to DNe and NLs \citep{2005Ramsay,2006Ramsay,2023Begari}.  \xmm\ and \swi\ observations of the outburst spectra  show
some X-ray suppression in flux and luminosity and cooling in hard X-ray temperatures during outburst (characteristics similar to DNe), 
but only by a factor of 1.5--2 times { \citep[][and references therein]{2012Ramsay,2019RiveraSandoval,2021Rivera}.} 
Moreover, the duration of active/outburst states is found longer
than DN and  is inconsistently large (5--10 times larger) compared to DIM outburst-duration predictions for orbital periods longer than about 30 min up to about 68 min (longest orbital period); cf. Figure 4 \citep{2021Rivera}.

\begin{figure}[H]
\begin{tabular}{ll}
\hspace{-0.25cm}
\includegraphics[height=2.1in,width=2.6in,angle=0]{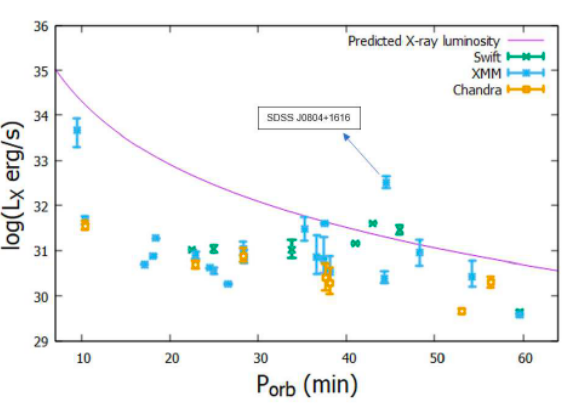} &
\includegraphics[height=2.1in,width=2.6in,angle=0]{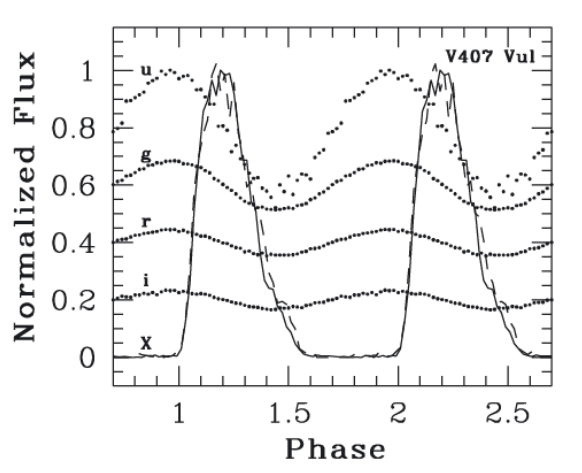} \\
\end{tabular}
\caption{On the left  is the X-ray luminosity distribution of AM CVns \citep{2023Begari}, Copyright 2023,  AAVSO, used with permission. On the right are the X-ray pulsations of V407 Vul detected with \cha\ along with optical variations~\citep{2009Wood}. \label{fig:li}
}
\end{figure}

The UV luminosity of these systems shows strong dependence on the 
orbital period due to hotter disks or higher accretion rates at shorter periods,
whereas the X-rays show no such dependence (true for most CVs as well) \citep{2007Ramsay,2010Solheim}.   
Archival X-ray data of \cha, \xmm, and \swi\  have been used to calculate the X-ray luminosity of several systems (28) to compare them with the orbital periods. Long-orbital-period, lower-bolometric-luminosity systems match model predictions in luminosity, but high-mass-accretion-rate systems, where optically thick boundary layers are expected, are sub-luminous in X-rays in most cases, as opposed to the expectations of  standard disk and BL theory, as clearly seen in the left-hand side of Figure~\ref{fig:li} \citep{2023Begari}. 

HM Cnc and V407 Vul  are characterized by ``on/off'' X-ray light curves that trail the optical light curves 
\citep[][]{2004Cropper,2007Barros} (see Figure~\ref{fig:li}, right-hand side).  The proposed model for these systems is the direct impact model with no disk involved \citep{2001Nelemans,2002Marsh,2008Dolence}. At sufficiently short orbital periods, the accretion stream can impact the surface of the WD directly. The accreted matter would thermalize below the photosphere with a temperature cool enough to emit soft X-rays. These two systems have been studied with \cha\ (LETG), indicating some line emission ($\sim$27~\AA, 31.6~\AA, and 46.4~\AA) and no C, N, O, and Ne lines. The spectrum is consistent with 65 $\pm$ 5 eV blackbody radiation for HM Cnc along with strong X-ray pulsations at the orbital period \citep{2003Israel,2008Strohmayer}. The \cha\ observation of V407 Vul also shows strong pulsations at the orbital period and a blackbody temperature of {72$^{+12}_{-8}$ eV} \citep{2008Ramsay,2004Strohmayer-c}. The two sources also show absorption features and enhanced abundances (e.g., Neon).

\section{WD Symbiotics}\label{sec:wdsim}

What powers symbiotic stars? Are they SN Ia progenitors? These are among the most common questions in the study of symbiotic stars. \cha\ and \xmm, together with other X-ray missions, have played a significant role in our quest for answers to those questions. Both instruments unveiled persistent emission from the residual burning white dwarfs in the Milky Way and galaxies in the Local Group, while in nearby galactic symbiotics, they also resolved spatially extended X-ray emission arising from collimated, jet-like structures. We summarize here some of those findings.


Symbiotics, binaries where a white dwarf or a neutron star accretes from a red giant companion in a wide orbit, have been identified as X-ray sources since GX~1+4 was observed with a balloon experiment by \cite{lewin71}. GX~1 + 4 is a member of the symbiotic X-ray binaries group, where the compact object is a neutron star. Being fainter X-ray sources (L$_x$ = 10$^{30}$--10$^{34}$ \lumcgs; \citep{2024Lima}), the first symbiotics with accreting white dwarfs were detected by the \Einstein\ Observatory in the early 1980s \citep{allen81}.  Since then, the current census of symbiotics detected in X-rays includes 65\footnote{\url{https://sirrah.troja.mff.cuni.cz/~merc/nodsv/utilities/x-rays.html}, { (accessed on 24 Feb., 2025)}} members \citep{merc19db} where 15 are NS binaries. While in symbiotic X-ray binaries, the X-ray emission is thought to arise mainly by Comptonization of photospheric radiation of the neutron star by a hot coronae, in the case of white dwarf symbiotics,  a variety of mechanisms  produce detectable X-ray emission.

High-luminosity, low-temperature, blackbody-type X-ray emission (known as ``supersoft'' X-ray emission) has been detected in a sample of white dwarf symbiotics even long after nova-type eruptions (see the following sections for the symbiotic novae in outburst). These ``long-lived'' supersoft sources were dubbed $\alpha$-type symbiotics in the first-proposed classification scheme introduced by \cite{muerset97} based on observations obtained with the \rosat\ mission. Other symbiotics showed X-ray emission with temperatures and luminosities of optically thin thermal plasma (bremmstrahlung) most likely arising in regions where strong shocks are produced. Such regions could be those where the winds from the white dwarf and the red giant collide. \citet{muerset97} called them $\beta$-types. Given the relatively narrow energy range covered by the \rosat\ mission, those symbiotics with energies $>$2.4 keV and spectra not compatible with those arising from a thermal plasma were classified as $\gamma$-types and were precisely the first members of the symbiotic X-ray binaries.

\subsection{Symbiotic Binaries in the Era of \xmm\ and \cha}
The advent of instruments with sensitivity in a broad 0.3-to-10 keV energy range, with high spectral and spatial resolution, \xmm\ and \cha, allowed pointed X-ray observations of known symbiotics, uncovering then-unknown features. 

Taking advantage of the high sensitivity (and effective area) of \xmm\ in the 0.3--1 keV range, the supersoft X-ray emission, related to either classical novae (see Section~\ref{sec:nova}) or symbiotic novae\footnote{A small fraction $\sim$10\% of symbiotic stars undergo very slow and long lasting nova outburst (several decades to a century) \citep{2024Munari}.} from known symbiotics such as AG~Dra \citep{agdra_xmm}, RR~Tel \citep{rrtel_xmm}, LIN~358, and SMC 3 \citep{smc3_lin358} have been revisited and precisely modeled. 
X-ray observations with \xmm, for the first time in a high state, of AG~Dra helped to uncover the cause of the anticorrelation between X-rays and UV/optical fluxes \citep{agdra_xmm,agdra_skopal}. Using the UV data from IUE, FUSE, and optical spectroscopy, \citet{agdra_skopal} proposed that the increase in the wind from the WD enhances the amount of particles in the nebula, which will afterwards contribute to the quenching of the supersoft X-ray emission by increasing the bound-free absorptions.  

The prototype of the class of symbiotic novae (outbursts that are thought to be due to the same physical mechanism as classical novae but last significantly longer), RR~Tel, was observed with \xmm\ in 2009. \citet{rrtel_xmm} found that the supersoft emission continues even after 65 years of its 1944 outburst, at a high luminosity of 10$^{37}$ \lumcgs. Moreover, the wide energy range of \xmm\ when compared with previous \rosat\ observations allowed the detection of the optically thin X-ray emission that arises from the shock of the WD and red giant winds; its temperature, once translated into velocity assuming strong shock conditions, is commensurate with the observed wind features in \hst\ UV spectra. 

The persistent nuclear burning phase is attributed to such a high mass accretion rate, where all the energy is {radiated as soon as it is produced}, without ever reaching the conditions of a thermonuclear runaway. This phase typically lasts only months to years after normal nova eruptions in cataclysmic variables (see this review), with the duration of the supersoft phase, at first order, being a function of the white dwarf mass. However, there is at least one symbiotic star that was observed in a nova outburst and showed indications of a longer nuclear burning phase than predicted, possibly 80 years. As noted by \citet{rrtel_xmm}, the WD mass in RR~Tel should be $\lesssim$0.65 \msun, in order to explain such a long burning phase in a postnova. Notably, \citet{rrtel_xmm} presented different ways of estimating the mass of the envelope, all of which produced values greater than a few 10$^{-5}$ \msun. If this material originates from the ejecta in a symbiotic nova outburst, it is significantly greater than the ejected mass in other nova eruptions in symbiotics. 

Outside the Milky Way, towards Local Group Galaxies, in which direction absorption is lower, \xmm\ and \cha\ have observed $\alpha$-type symbiotics in detail and repeatedly. \citet{smc3_lin358} and later \citet{2024RNAAS...8..127B} reported observations of LIN~358 and SMC~3, two symbiotics undergoing significant nuclear burning, but without any known outburst in the last century (in contrast to the case of RR~Tel). These two objects seem to be accreting at an unusually high rate of a few 10$^{-7}$ \msun\ yr$^{-1}$, precisely the rate predicted to be necessary for stable nuclear burning \citep{Fujimoto1982}. By modeling multi-wavelength observations of Lin~358, other researchers \cite{linb358_kuuttila,2022AJ....164..145S} concluded that it could be accreting at an even higher accretion rate, possibly at super-Eddington rates, which certainly calls for further follow-up observations.

The wide field of view of \xmm\ has been crucial to the observation of the X-ray population of external galaxies and in this sense has allowed the discovery of candidate symbiotics of all X-ray types. \citet{saeedi_dracoC1} reported the observations of Draco~C1, an $\alpha$-type symbiotic, which also accretes at a very high rate. During these observations, two new symbiotics of the $\beta$-type class were detected for the first time in an external galaxy \citep{saeedi_draco}. Later, \citet{saeedi_wilman1,saeedi_sculptor} reported their finding of one and three symbiotics in the galaxies Wilman~1 and Sculptor, respectively.

\begin{figure}[H]
\begin{adjustwidth}{-\extralength}{-4.4cm}
\centering
\includegraphics[scale=0.75]{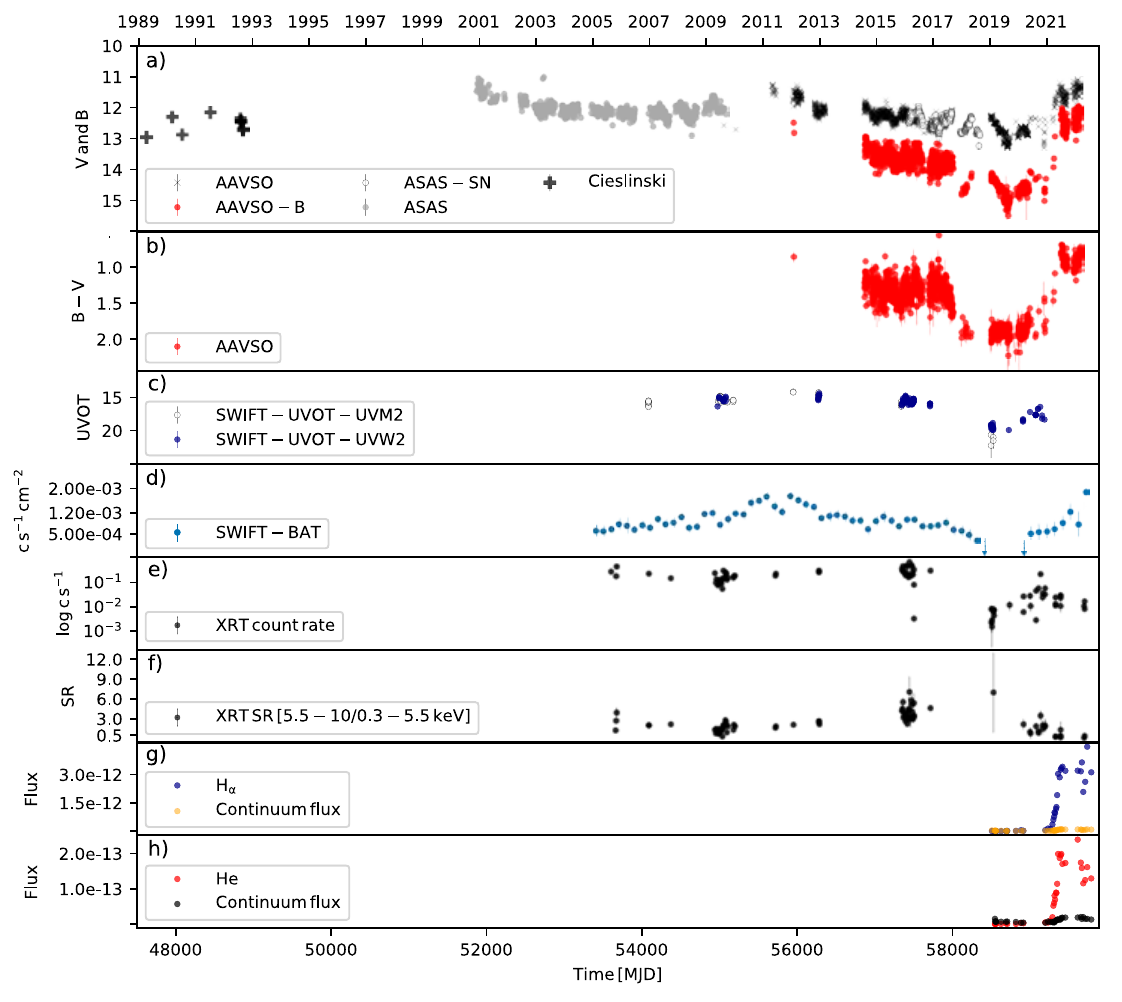}
\end{adjustwidth}
\caption{Multi-wavelength time evolution of the symbiotic binary RT~Cru. This is figure 2 from \citep{pujol2023}. \label{fig1_jl}}

\end{figure}

The launch of wide-field, hard X-ray sensitive instruments such as \swi/BAT and \integral\ provided the first evidence of a hidden population of very hard X-ray-emitting symbiotics.  \citet{integralRTCru} reported the detection of a new hard X-ray source that was later identified as the already-known symbiotic RT~Cru \citep{XRTRTCru,BATRTcru}. The \swi/XRT spectrum did not have enough counts to allow a constrained modeling, and a subsequent 50 ks \cha/HETG observation provided the first high-signal-to-noise X-ray spectrum \citep{chandraRTCru}. Even the zero-th-order spectrum of this \cha\ observation showed the presence of emission lines in the Fe K$\alpha$ region arising from a thermal plasma, suggesting that the X-ray emission in RT~Cru was driven by a different mechanism than that in symbiotic X-ray binaries. The spectrum was modeled with a multi-temperature thermal plasma where the emissivity is inverse to the bolometric luminosity \cite{mkcflow}, i.e., a {\em cooling flow}. This model has been used to fit the X-ray spectra of magnetic cataclysmic variables (IPs) and dwarf novae in quiescence \citep{2005Pandel,2003Mukai}, where, after a shock, the X-ray plasma cools off until it reaches the white dwarf surface. The temperature at the shock region (kT$_{max}$ parameter in the \texttt{mkcflow} model within XSPEC\footnote{\url{https://heasarc.gsfc.nasa.gov/docs/xanadu/xspec/}} is commensurate with the WD mass, which dictates the speed at which the strong shock is produced. In their interpretation, the authors of \citep{chandraRTCru} found that RT~Cru might host a massive WD, with $M_{WD} \gtrsim$ 1.3 \msun. Follow-up \cha\ and \xmm\ observations have proven the extremely variable nature of this source. \citet{ash2021,ash2024} found that the observed variability in short timescales is most likely due to fast changes in the amount of absorbing material, while on the other extreme, \citet{rtcru2018,pujol2023} found that the long-term variability is due to changes in the accretion rate towards the white dwarf as it travels in a wide orbit around its red giant companion (Figure \ref{fig1_jl}).

\begin{figure}[H]
\begin{adjustwidth}{-\extralength}{-4.cm}
\centering
\includegraphics[scale=1.12]{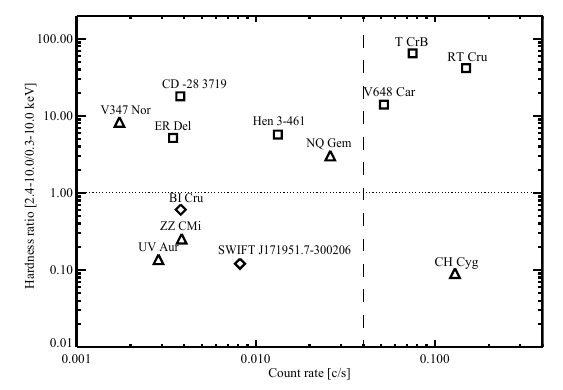}
\end{adjustwidth}
\caption{Hardness ratio vs. count rate for the hard X-ray WD symbiotics observed with \swi. The hard X-ray symbiotics detected with Swift are located to the right of the vertical dashed line. The plot shows that $\delta$-type sources are located above the dotted horizontal line, where hard X-ray photons dominate the spectrum; $\beta$-type sources, on the other hand, show a hardness ratio of less than 1; and $\beta$/$\delta$-type sources are located at both above and below the hardness ratio of 1. 
The plot shows that $\delta$-type sources (squares) have a hardness ratio of more than 1 (above the dotted horizontal line); $\beta$-type sources (diamonds) have hardness ratio of less than 1; and $\beta/\delta$-type sources (triangles) are located above and below the hardness "ratio = 1" line. This plot is Figure 5 in \citep{luna_2013}, reproduced with permission by ESO. \label{fig2_jl}}
\end{figure}

The number of white dwarf symbiotics with hard X-ray emission has grown since then, including now SU~Lyn \citep{sulyn2018}, V648~Car \citep{ss73_eze}, T~CrB \citep{tcrb_suzaku}, and CH~Cyg \citep{chcyg_suzaku}. All of them show hard X-ray emission arising from an optically thin thermal plasma at a high temperature \citep{kennea_2009,integral_review} that is subject to significant absorption. Subsequently, to classify those symbiotics with hard (E $\gtrsim$ 2.4 keV) X-ray emission of thermal origin and affected by significant absorption, \citet{luna_2013} added the category of $\delta$ type in the classification of X-ray emission that was first introduced by \cite{muerset97} (see Figure \ref{fig2_jl}). The temperature of the hard X-ray emission (10--50 keV \citep{kennea_2009}) points to a shock origin with matter reaching speeds of a few thousands km s$^{-1}$, and its most plausible origin is thought to be the accretion disk boundary layer around a massive white dwarf. In analogy to dwarf novae in quiescence, this emission would arise from the boundary layer once the accretion rate is below a few 10$^{-9}$ \msun\ yr$^{-1}$ \citep{1993Narayan}, where the plasma would have a low optical depth and is not efficiently cooling. Given that the temperature would depend on the shock speed that depends on the gravitational potential of the white dwarf, fitting the X-ray spectrum directly provides a {\it lower limit} on the mass of the white dwarf. We emphasize that this measurement provides a lower limit on the mass because the presence of additional cooling mechanism/s, such as Comptonization, would first dominate the shock process, and thus, once the plasma starts radiating thermally in X-rays, it has already slowed down. However, no other cooling mechanism, aside from optically thermal emission, has been identified in these sources, and in all cases but CH~Cyg, there seems to exist a white dwarf with a mass $\gtrsim$1.1 \msun.

In a handful of nearby symbiotics, their nebulae and jets can reach angular sizes of a few arcsec, which can be resolved with \cha. Two symbiotics stand out in this regard, R~Aqr and CH~Cyg. \citet{chcyg_jeno} reported the detection with \cha\ of extended X-ray emission that matches the location of the previously detected radio \citep{vla_chcyg} and optical \cite{hst_chcyg} structure. It is now accepted that this structure is a precessing jet, a conclusion derived after follow-up observations obtained with \cha, \hst, and \vla\ by \cite{chandra_chcyg2}. This conclusion was drawn thanks to innovative imaging analysis techniques that allowed the exploration of the \cha\ spatial resolution well below its nominal value of 0$^{\prime\prime}$.492. 

\begin{figure}[H]
\begin{adjustwidth}{-\extralength}{-4.5cm}
\centering
\includegraphics[scale=0.67]{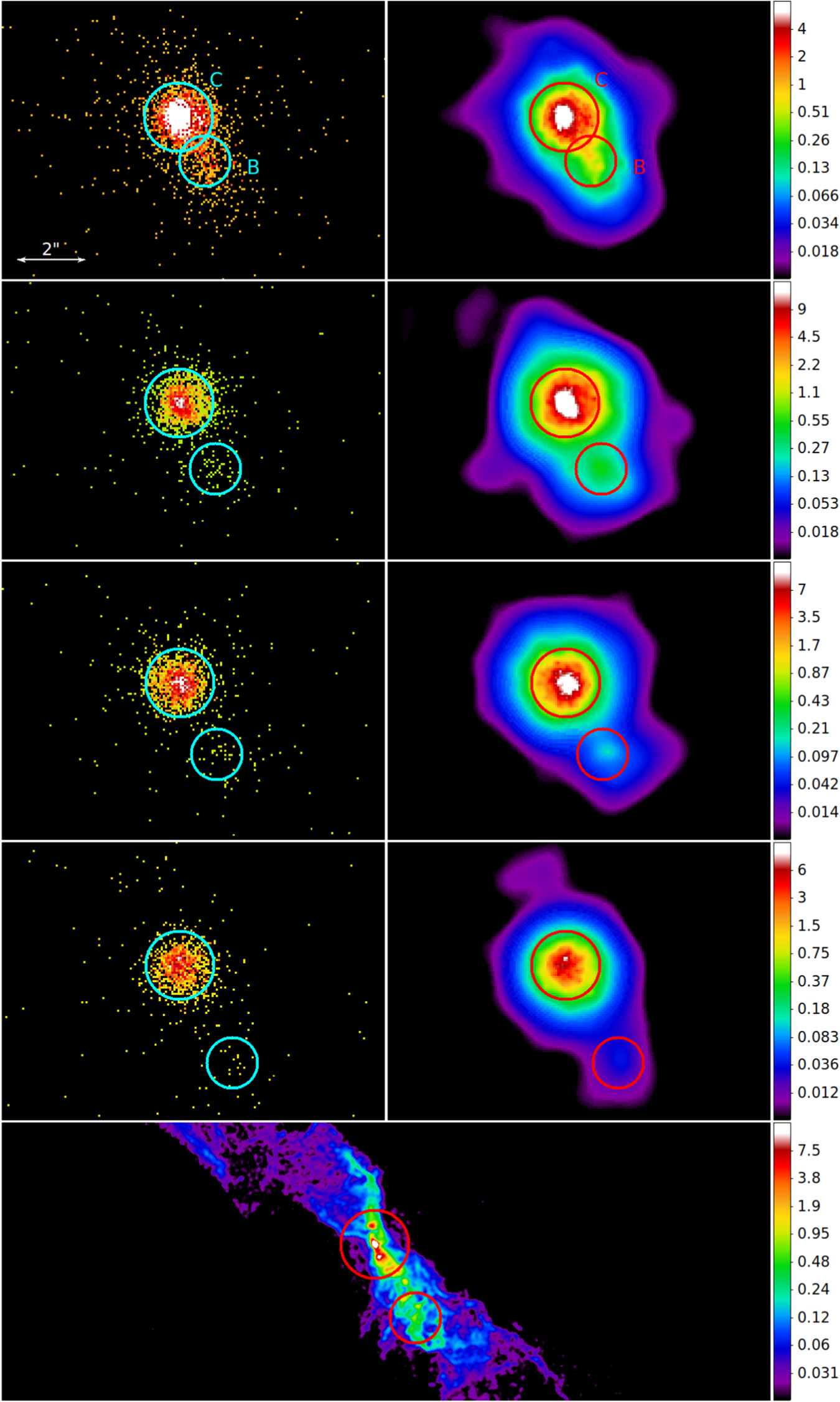}
\end{adjustwidth}
\caption{ Soft band (0.5--2.0 keV) images of R~Aqr obtained with \cha\ on 2017, 2020, 2021, and 2022 (from top to bottom). On the left is the raw image, and on the right are the adaptively smoothed images. The circles show the central region (C) and blob (B) regions. The evolution of the blob in the jet is remarkable. {The last row shows an HST image obtained in 2021 with the F373N, $\lambda$ 3730 \AA, and [O II] line filter. All Chandra images are in units of counts. The HST image is in units of electron/s.} { This image is Figure 7 from \citep{raqr_2024}, reproduced under CC-BY 4.0 license AAS}. \label{fig3_jl}}
\end{figure}

\citet{raqr_kellogg} reported the first detection in X-rays of the jet-like structures in R~Aqr (see Figure~\ref{fig3_jl}) . These jets, previously known in other wavelengths, were resolvable thanks to the unprecedented and so-far-unmatched \cha\ spatial resolution. Both NE and SW jets emit soft (with energies of less than about 1 keV) X-rays, while the hardest X-rays detectable with \cha\ originate in the central source. The evolution of the soft X-ray emission in a yearly timescale has been tracked with \cha, \xmm, and \swi\ by \citet{raqr_2024} as displayed in Figure~\ref{fig3_jl}. Most likely its increase in 2020 was related to the periastron passage, when the accretion onto the WD might have increased. 

Remarkably, the $\beta$/$\delta$-type X-ray spectrum is a characteristic of most symbiotics with collimated/jet structures. Although the X-ray emission has been resolved only in CH~Cyg and R~Aqr, sources such as MWC~560 \citep{mwc560_lucy}, Hen~3-1341 \citep{stute_1341}, and V347~Nor \citep{luna_2013} show X-ray spectra consistent with emission from at least two optically thin thermal plasmas, with temperatures of a few million and tens of millions of degrees. The low-temperature plasma, in analogy to those systems with spatially resolved X-ray emission, could arise from shocks within the jets, while the high-temperature emission might be related to the accretion process onto the WD.


\section{Novae in Outburst}\label{sec:nova}

Novae in outburst are amazing astrophysical laboratories. The outburst is initiated by thermonuclear CNO burning of hydrogen, ignited at the bottom of a shell accreted on a white dwarf (WD) from a binary companion. At some point, burning becomes explosive because of electron degeneracy. In the ensuing thermonuclear runaway, novae increase in optical luminosity by 8 to 17 magnitudes within hours, and eventually eject accreted material in a very fast wind.

{The bolometric luminosity of novae remains at the Eddington level} ($\simeq$$1.2 \times 10^{38}$ erg s$^{-1}$ for a 1 M$_\odot$ star), for a period ranging from a few days to a few years, emitting copious flux at all wavelengths, from radio to gamma rays, up to the range of Cherenkov telescopes~\citep{Franck2018,Cheung2022,Acciari2022,HESS2022}. Novae in outburst are known as luminous X-ray sources since the 80s with \exo\ \citep{Ogelman1984}, and \rosat\ proved that
the X-ray light curve and spectra are as important and interesting as the optical, UV, and { IR ones, e.g., \citet{Balman1998}.}  Novae become in fact some of the brightest sources in the X-ray sky, in two phases of different phenomena, producing rich grating spectra: the shock phase and the luminous supersoft X-ray phase (SSS). These spectra will be an important legacy of \cha\ and \xmm, also because a very wide range of X-ray atomic transitions are detected, contributing  to the atomic databases.

There is another reason for which the \cha\ and \xmm\ observations have revealed unexplored physical phenomena. With the aim of using the gratings, long exposures were required, revealing aperiodic and periodic variability. Some phenomena of periodic variability have revealed the presence of strongly magnetized white dwarfs, e.g.,~\citep[][]{2021Drake,2024Orio}.

\subsection{The High-Resolution Spectra: The Shocks Phase}

The location and nature of the shocks, as well as the overall dynamics of the ejecta, are still being explored  (e.g., \citep{Diesing2022}).  On average, every 4 years, a Galactic nova is luminous enough in this phase to allow exploring the ejection dynamics,  the chemical enrichment, and the connection with the mechanisms producing radio and gamma-ray flux. The absorbed peak X-ray luminosity of most nova shells is close to  10$^{34}$ erg s$^{-1}$, but it is two or three orders of magnitude higher in novae occurring in symbiotics, systems with a red giant or AGB companion.  Powerful shocks originate if the thick wind of the giant is impacted by the blast of the nova \cite{Nelson2008, Ness2009, Drake2016, Orlando2009, Orlando2012, Orlando2017, Orio2020, Orio2022}, but can also arise very close to  the red giant, in the impact with a non-disrupted accretion disk, or in  the initial, fast bipolar outflow (observed, for instance, in RS~Oph \cite{Munari2022}). Symbiotic novae in the Galaxy occur only once every $\approx$7 years, but  the relatively nearby symbiotic nova (d$\simeq$800--900 pc) T~CrB is likely to have its third recurrent outburst quite soon \citep{2024Shara}.

The plasma temperature is proportional to the shock velocity, given by kT$\simeq$1.2 keV \mbox{(v/(1000 km/s)}$^2$; typical velocities are in the 1000--6000 km/s range, and {the peak temperature in  the shocks can be determined only with high-resolution X-ray spectra}. Usually, thermal emission of a plasma in collisional ionization equilibrium (CIE) has been assumed, but recently, the recurrent symbiotic nova RS Oph, observed with \nicer\ in outburst in 2021, has shown non-equilibrium effects \citep{Orio2023, Islam2024} at least in the first week. The gas was at a peak temperature of 20--27 keV (so the plasma should have been almost fully ionized); however, the flux in the He-like lines was larger than in the H-like ones (in particular, the Fe XXV lines were much stronger than the Fe XXVI \citep{Orio2023, Islam2024}).  Models of thermal plasma in non-equilibrium ionization were adopted to derive the physical parameters, in which the electron density of the  emitting plasma is inferred from a fit parameter, the upper limit to the ionization timescale.

\begin{figure}[H]
\begin{adjustwidth}{-\extralength}{-4.5cm}
\centering
\includegraphics[width=13.4cm,height=7.6cm]{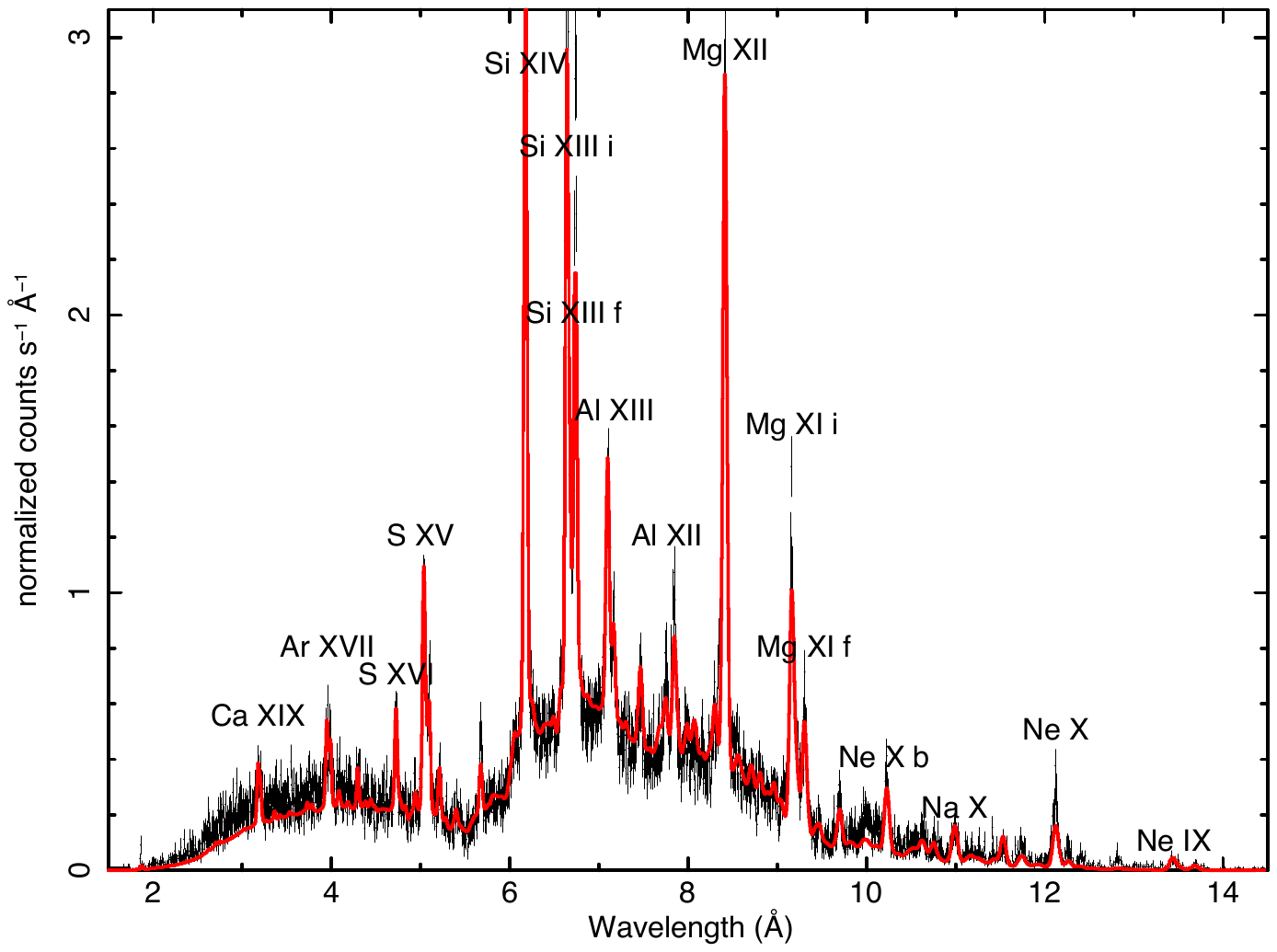}
\end{adjustwidth}
\caption{ RS Oph \cha\ HEG spectrum on day 18 of the 2021 outburst. Figure is obtained from \cite{Orio2022}. The composite-fit with two APEC plasma components, at 1 keV and 0.22 keV is plotted in red color. The measured flux decreased with a factor of 3 from day 2, but the estimated absolute flux had decreased by a factor of 15 times \cite{Orio2023}. All detected emission lines are labeled, reproduced under CC-BY 4.0 license AAS. \label{nov:1}}
\end{figure}

Unfortunately, in 2021, RS Oph could only be observed with the HETG on day 18 after the optical maximum and with the RGS on day 21; by that time, the spectrum could be fitted well with two components of plasma in CIE (see Figure~\ref{nov:1}). Another recurrent symbiotic nova, closer to us, T CrB, is expected to have an outburst soon \citep{2023Ilkiew}. The \cha\ HETG in the first week, and soon after the RGS, will provide an exact estimate of the peak flux and temperature. We note that the absolute flux estimate is critically dependent on the absorbing column  density. In the initial stage, the connection with the gamma-ray emission can be throughly explored. So far, only two novae have been semi-simultaneously observed and detected in gamma-rays and X-rays, RS Oph \citep{Luna2021} and N Sco 2023 \citep{Sokolovsky2023}, allowing the connection with the gamma-ray emission to be explored.

As the gas quickly cools and reaches equilibrium, {the X-ray spectra always become very rich in emission lines due to many transitions} of H-like, He-like, and other ions, including L-shell iron lines, critically constraining models of the dynamics and geometry of the expanding and colliding ejecta, such as those developed by \citep{Orlando2009, Orlando2012, Orlando2017}. The density diagnostics and line profiles late in the outburst \citep{Tofflemire2013, Peretz2016, Orio2020} indicate {\it an extremely clumpy wind}, which perhaps gives rise to the cooler ``knots'' observed  optically at late times (e.g., \citep{Shara2015}).

The majority of novae do not have evolved donors with thick wind that mixes with the ejecta; thus, the shocks originate in material ejected from the WD at different velocities (colliding winds) or collisions between the ejecta and the accretion disk or likely the circumbinary gas. Abundances revealing the WD nature, the temperature, and the electron density in the shocked outflow material can only be derived from the high-resolution spectra, thanks to the atomic transitions in the soft X-ray range. Intermediate atomic number elements, mostly S, Ar, Cl, Ca, and Al, are overabundant only on oxygen--neon (ONe) WDs, where Ne-Na and Mg-Al cycles, which do not occur
on carbon--oxygen (CO) WDs, operate in addition to the CNO cycle \citep{Jose1998, Starrfield2009, Kelly2013}; there is also a Mg-rich  superficial layer. The models predict that in recurrent novae (RN, novae with high-mass WD and outbursts repeated every 1--50 years  with ejecta mass lower than CN), the WD mass grows despite the cycles  of mass loss in the eruptions. Thus, they are candidate SN Ia progenitors, if they host CO WDs (M$_{WD}$ = 0.4 \msun--1.05 \msun~\citep{2018Lauffer}). ONe WDs---otherwise rare, but frequent in novae because of selection effects---do not  ignite carbon, but they collapse into neutron stars if they grow in mass  (M$_{WD}$ $>$ 1.05 \msun~\citep{2007Siess}). On the other hand, they  are very important in Galactic chemical evolution. For instance, rates of nova aluminum production
critically constrain the rate of SNe II  (see \citep{Iliadis2011, Bennett2013, Peretz2016}),  and the radioactive $^{26}$Al isotope (producing a well-observed, strong gamma-ray line at 1809 keV) causes water evaporation in proto-planetary disks \citep{1999Srinivasan}.

\begin{figure}[H]
\includegraphics[width=13cm,height=6.5cm]{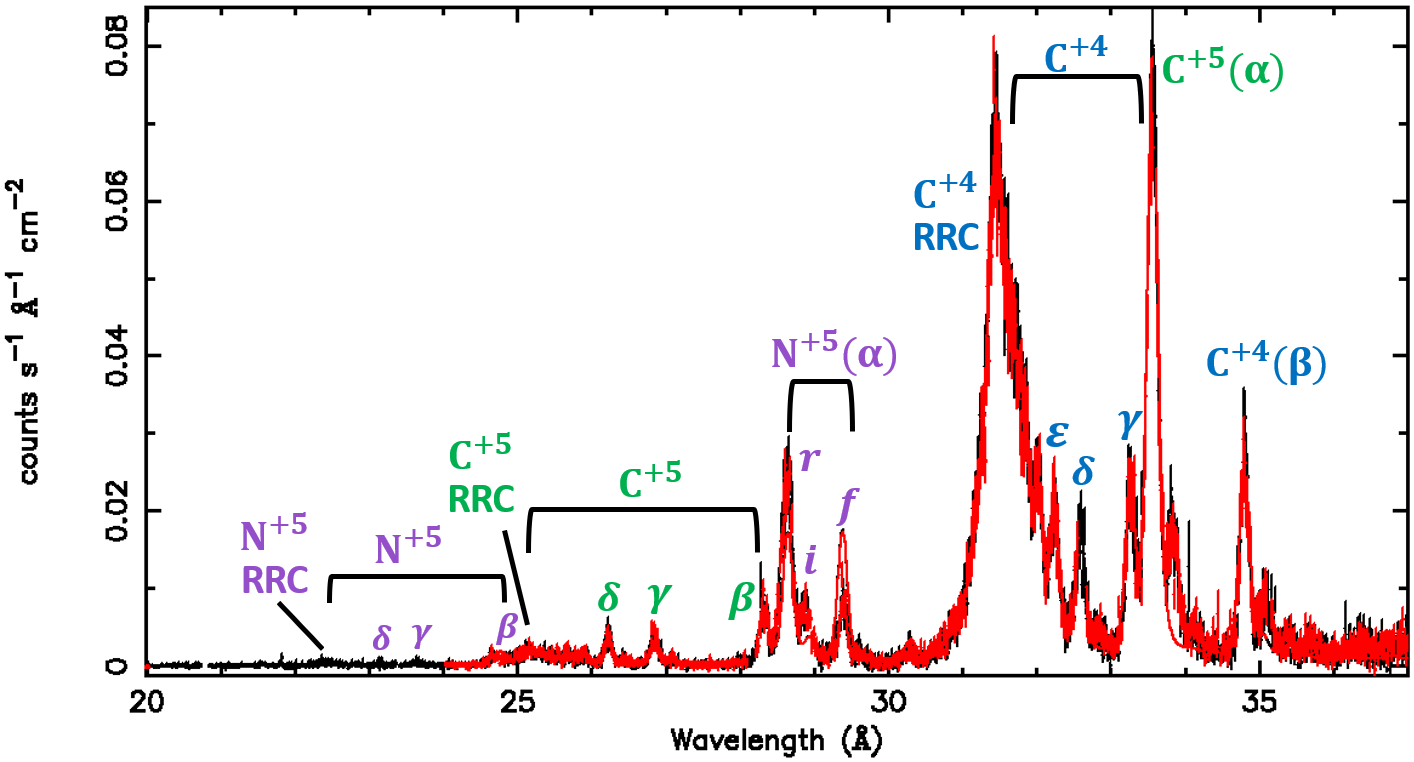}
\caption{From \citet{Mitrani2024}: \xmm\ RGS1 and RGS2 grating spectra of YZ Ret 77 days after the outburst (30 ks exposure), including rarely observed radiation recombination continua (all the features of the spectrum are as labeled, reproduced under CC-BY 4.0 license AAS.\label{nov:2}}
\end{figure}

Finally, {shocks seem to persist for quite a long time, while the emission cools}. \citet{Contini1995} inferred that the optical and UV spectra of RS Oph in 1985 still showed clear signatures of shocked
material, and in fact, \citet{Nelson2008} and \citet{Ness2009} 
presented spectra of  X-ray emission lines months after the 2006 outburst, after the white dwarf spectrum of the SSS had faded completely.  Only the X-ray gratings distinguish clearly between spectra of the SSS and those due to the much-cooled shocked plasma and the interaction of the outflow with circumstellar material. For instance, the spectra of YZ Ret (Nova Ret 2020) were ``supersoft'', but lacked a atmospheric continuum; they
were extremely rich only in emission lines, revealing interesting atomic physics, {including narrow radiation recombination continua and charge exchange phenomena} (see Figure~\ref{nov:2} and \citet{Mitrani2024}).
Thus, novae can be amazing templates to explore and learn rarely observed atomic physics.

\subsection{The \xmm\ RGS, \cha\ LETG Grating Spectra: The Central White Dwarf as a Luminous Supersoft X-ray Source}
 
The post-maximum optical decline of  a nova is accompanied by a shift in the wavelength of maximum energy towards shorter wavelengths, in a constant bolometric luminosity phase at the Eddington level,
as the WD photosphere shrinks back to a pre-outburst radius, while thermonuclear burning still occurs near the surface. {The WD becomes a supersoft X-ray source (SSS) with an effective temperature of T$_{\rm eff} >$ 250,000 K}, exactly as predicted by the models \cite{Starrfield2012, Wolf2013}. The SSS may be easily absorbed with high column density, but in most cases, the ejecta become optically thin to supersoft X-rays. {The \cha\ LETG samples the softest portion of the nova WD spectrum, allowing accurate column density measurements}.

{ By modeling the LETG spectra, abundances, effective temperature, WD effective gravity, composition, and  mass are derived}. The data have been fitted with two classes of models: of non-local thermodynamic equilibrium static atmospheric models for hydrogen-burning WDs (TMAP, \cite{Rauch2010}), and ``simpler'' photoionization models, like PION in SPEX~\citep{Ness2011,Pinto2012,2017Balman}. The atmospheric models include the effective gravity as a parameter; they yield a very good match with the depth and strength of the absorption features, and often also with the continuum shape and absorption edges.  However, when the SSS emerges, blueshifts corresponding to velocity up to 3000 km~s$^{-1}$ are observed,  implying that the WD atmosphere is still emitting a wind, even when other diagnostics (and the current models) seem to indicate the end of the outflow.  The photoionization codes allow us to account for the blueshift velocity and estimate the mass lost. They also include more atomic transitions, and are flexible to explore a wide range of abundances \citep{Orio2021, Ness2022}.

The nature of the WD can also be assessed in this phase: carbon absorption edges reveal CO WDs, because the core material is eroded and mixed in the envelope. Carbon is depleted in the CNO cycle, but it is newly dredged up and is much more abundant, with deeper edges, in a CO WD than in an ONe one.

\begin{figure}[H]
\includegraphics[height=7.5cm,width=13cm]{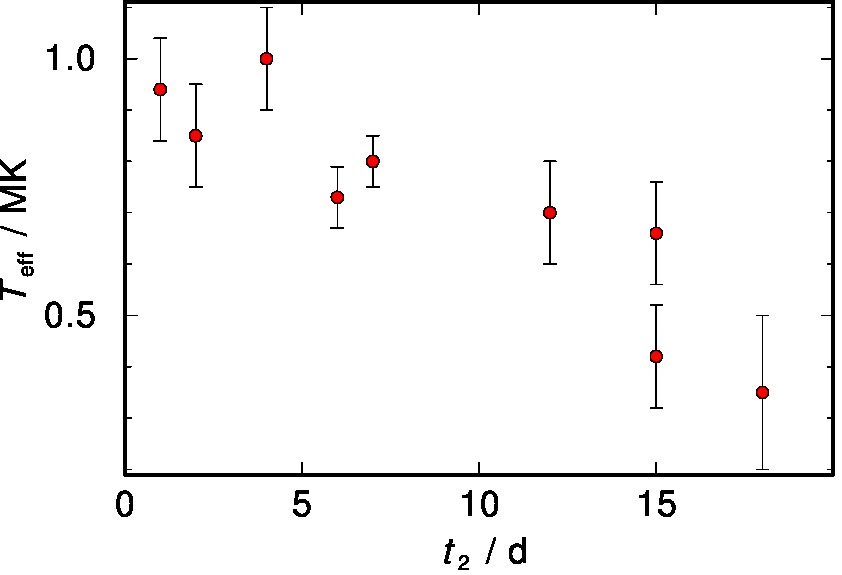}
\caption{ 
T$_{\rm eff}$ values from the literature  (including  broadband data), showing the almost linearly inverse correlation with $t_2$, time for decay by 2 mag in optical. Both parameters are predicted to be diagnostics of $M_{\rm WD}$, with spread mostly due to the role of the mass transfer rate. From left to right on the t$_2$ axis: U Sco \citep{Orio2013}, HV Cet  \citep{Beardmore2012}, V2491 Cyg \citep{Ness2011},V4743 Sgr \citep{Rauch2010},  RS Oph \citep{Nelson2008}, V5116 Sgr \citep{Sala2017}, V1974 Cyg (above) \citep{Balman1998}, T Pyx 2011 (below) \citep{Tofflemire2013}, and N LMC 1995 \citep{Orio1999}.
\label{fig:t_vs_m}
}

\end{figure}

\begin{figure}[H]

\includegraphics[width=14cm,height=8.0cm]{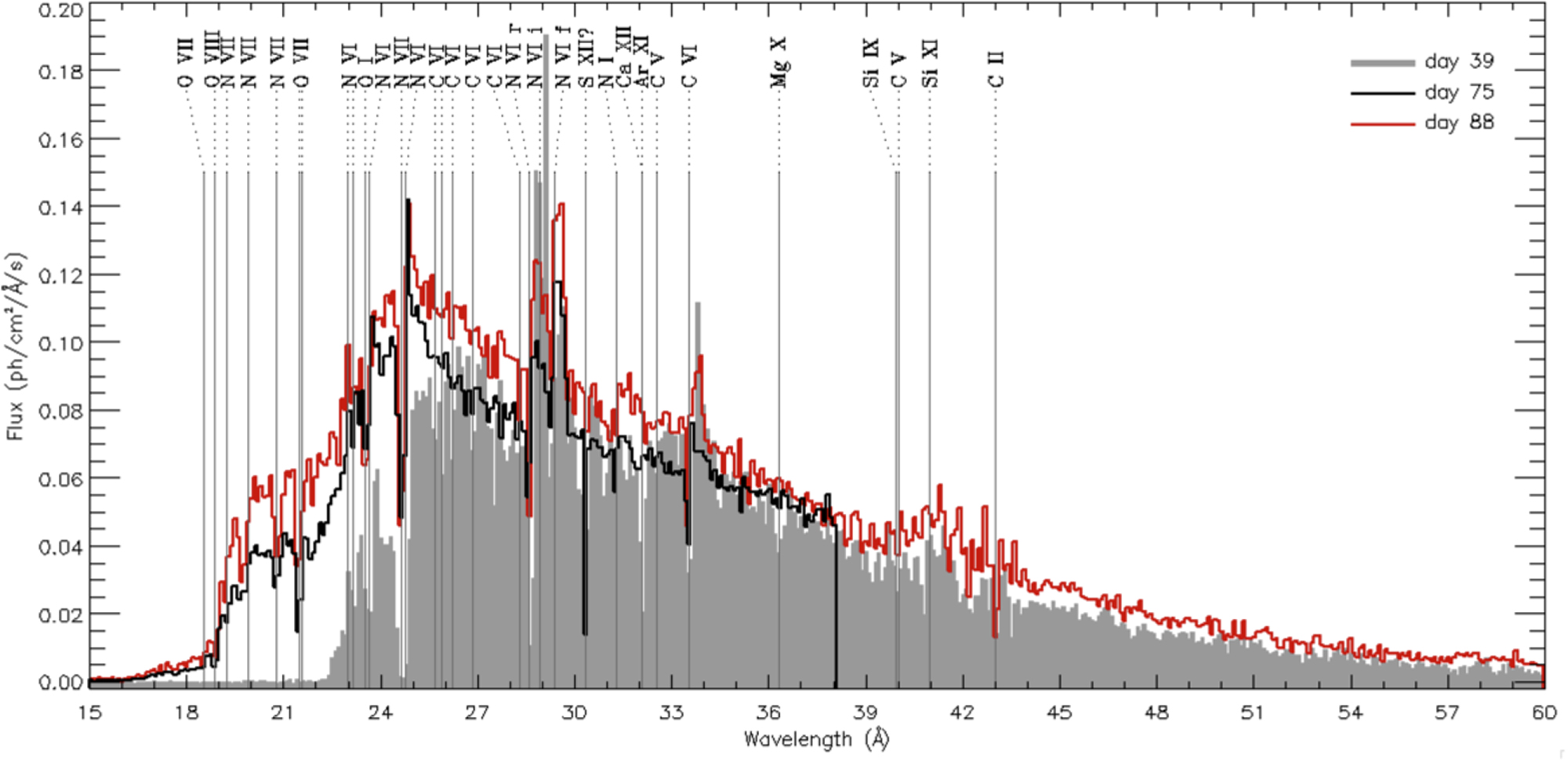}
\caption{\cha\ LETG (red and gray) and RGS (black) X-ray spectra of Nova SMC 2016a on days 39, 75, and 88 of the outburst showing the SSS evolution (from \citet{Orio2018}, reproduced with permission by AAS). The spectral lines---except for the ISM ones---are blueshifted by 1800 km/s.\label{nov:3}}

\end{figure}

The models are constrained by correlating the  X-ray spectral parameters with  those derived from other wavelengths. Several  predicted relationships between physical parameters of the outburst can be tested, but the parameter space can be analyzed only by examining  a number of different novae; moreover, each observation is a snapshot in the evolution. The peak T$_{\rm eff}$  of the hydrogen-burning WD is predicted by the theory to be, practically, a proxy for the WD mass. Figure~\ref{fig:t_vs_m} shows the correlation so far obtained for T$_{\rm eff}$  versus  t$_2$, the time for a decline by two magnitudes in optical, another parameter that, in the models, is also dependent on WD mass (see \cite{Wolf2013}).

Finally, the SSS continuum of novae illuminates the interstellar medium at the inner-shell transition energies of low-Z elements.  X-ray grating spectra are measured even for novae in the Magellanic Clouds (see Figure~\ref{nov:3} and \cite{Orio2018, Orio2021}), which shine as lamposts, allowing {analysis of the local ISM absorption lines and edges, including  benchmarking  atomic data}. Only the Chandra LETG covers the important range longward of 40 \AA. Carbon lines have been identified and modeled
with Galactic novae  \cite{Gatuzz2018}, and initial results were obtained from an SMC nova by \cite{Orio2018}. Abundances, ionization fraction, and column densities of the local ISM along the line of sight of the Magellanic Clouds can be studied through these spectra.


\subsection{The Surprising Light Curves}

\begin{figure}[H]
\begin{adjustwidth}{-\extralength}{-4.5cm}
\centering
\includegraphics[height=9cm,width=9cm]{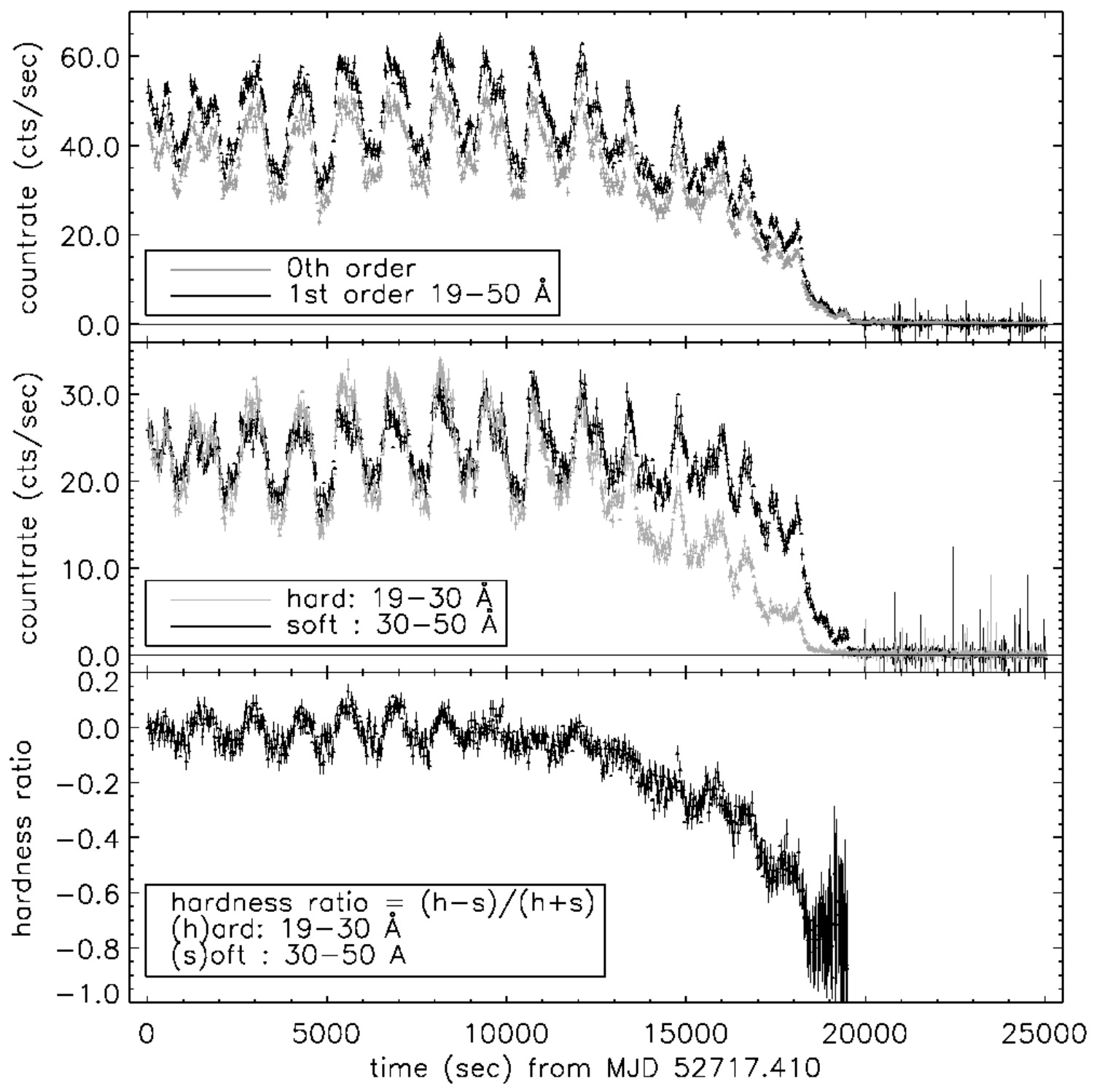}
\end{adjustwidth}
\caption{Chandra light curves of V4743~Sgr, and Nova Sgr 2002. This figure is obtained from \citep{Ness2003}, reproduced with permission by AAS.\label{nov:5}}

\end{figure}

The long exposure times required for the grating observation have revealed {variability phenomena during the SSS phase}, which were quite unexpected. Both the aperiodic and the periodic variability already appeared in the first two observations performed  with the \cha\ LETG grating \citep{Drake2003, Ness2003}, as shown in Figure~\ref{nov:5} for V4743~Sgr. The ``disappearance of the SSS'' in this nova was extremely intriguing, and the only possible explanation was that new material was ejected by the nova and/or an opaque clump absorbed the flux of the SSS. In any case, several exposures of other novae in the SSS phase revealed intense irregular variability; see \citep[][]{Ness2011, Ness2012, Orio2013, Orio2023, Ness2023}. In the case of RS Oph, this was attributed to multi-ionization photoelectric absorption zones, representing different lines of sight for absorption~\citep{2011Osborne}. Altogether, these phenomena paint a picture of ``complicated'' nova outflows, rather different from a constant mass loss rate in a uniform wind.

The first nova observed with the LETG grating revealed a modulation with a period of $\approx$42.7 min \citep{Drake2003}. The second one, V4743~Sgr, showed a 22 min (1325 s) period \citep{Ness2003}, which was well studied in a series of following papers \citep{Leibowitz2006, Dobrotka2017}, well visible in Figure~\ref{nov:5}. In the case of V4743 Sgr, this was found to be the rotational period, observed also in quiescence~\citep{Zemko2016, Zemko2018}. While the periodicity in quiescence is interpreted as due to the rotation period of an intermediate polar (IP), in outburst, the amplitude of the modulation with respect to the huge supersoft luminosity cannot be explained by ongoing accretion. It seems that the burning under the polar caps of this ``magnetic nova'' was occurring in a layer closer to the surface, and the reason is still being explored. In any case, this phenomenon of variability was discovered during the LETG exposures of several other outbursts \citep{Drake2003, Ness2011, Peretz2016, Aydi2018, 2021Drake} and is generally interpreted as revealing the IP nature of these novae. Highly magnetized white dwarfs in novae may thus be more common than previously thought.

The SSS in novae often also shows {shorter periods, on the order of a minute or less}, and although this shorter periodicity is often revealed in short \swi\ and \nicer\ exposures \citep{Orio2023}, it is also quite well studied with the LETG and the HRC camera of \cha\ (see \citep[][]{Ness2007,Orio2021,Pei2021}), which have often followed the evolution of the period over timescales of many hours. The nature of this period is still debated. The fact that the period, especially early in the outburst, is not constant but has a sort of irregular drift (see \citep[][]{Orio2023}) seemed difficult to reconcile with the white dwarf's spin. A likely explanation is thought to be g-mode pulsations due to the burning; however, the observed period seems to be too long with respect to the models \citep{Wolf2018}, so the root cause is still the subject of investigations. {Recent observations of V1716 Sco (Nova Sco 2023) show that the apparent period drift is an artifact of varying amplitude of the pulsations, and this could be assessed only with long, uninterrupted \cha \ exposures (Worley et al. 2025, preprint). 
}

\subsection{The Return to Quiescence}

Because accretion phenomena are best studied in X-rays wavelengths, \cha\ and \xmm\ have also proved to be precious in following the return to quiescence and the onset of accretion in novae of different types. The theoretical models predict that {two fundamental parameters determining the evolution and outburst characteristics of novae are the mass accretion, $\dot{M}$, 
and the white dwarf mass}. As already outlined in Section~\ref{sec:cv}\ of this review paper, X-ray observations give the best opportunity to estimate these important parameters. Thus, much work has also been carried out in following novae whose physical parameters were well studied in outburst as they settled in quiescent accretion. Although some of these novae are old and were only known as optical sources, like V603~Aql (N Aql 1918) \citep{2005Mukai} and CP~Pup (N Pup 1942) \citep{Mason2013}, other novae observed with \cha\ and \xmm, even without the gratings, already belonged to the ``X-ray era'', like V4743~Sgr \citep{Zemko2015} and V407~Lup \citep{2024Orio}. Novae with typical periods of minutes observed in the SSS phase display modulations with the same periods in quiescence, revealing their IP nature \citep{Zemko2015, Zemko2016,2024Orio}. An interesting characteristic of these novae is a low-luminosity supersoft flux that appears to remain detectable for a few years, as if the SSS was shrinking instead of just cooling 
(see especially the case of V4743 Sgr, \citealt{Zemko2016}, where the soft flux eventually~faded).


\section{Final Remarks}

Studying AWDs involves understanding accretion physics, the accretion disks and the flow structure, winds, jet formation, thermonuclear explosions, and nucleosynthesis. 
There are numerous state changes and transient events including but not limited to novae.

AWD science is intricately linked to the final phases of stellar evolution and other binaries with compact objects, as well as to young stellar objects, active galactic nuclei, quasars, and remnants of supernovae. Such connections are key to a comprehensive and correct understanding of common astrophysical phenomena. AWDs are also important for evolutionary and cosmological studies,  as possible single-degenerate progenitors of SNe Ia.

In this paper, we have attempted to summarize the rich panorama of  scientific results obtained {on CVs and related systems (i.e., AWDs)} by the \cha\ and \xmm\ Observatories in their 25-year mission time-span. The characteristics of their onboard instrument---their sensitivity
and broad spectral resolution capacity, along with superb spatial resolution for \cha\ (appreciable spatial resolution for \xmm)---have been crucial in finding/confirming new AWD systems aiding population studies in the Galaxy, Globular Clusters, and beyond.
\cha\ and \xmm\ have immensely contributed to our knowledge and understanding of accretion of all types, transient states (changes of such states), disk outbursts, and explosive nuclear burning. Such studies have contributed to theoretical impacts and new models for these phenomena. The results of such studies have bridged the gap between AWDs and other compact object binaries, AGN, and supernovae.  

We would like to underline the important role of the high-spectral-resolution gratings. All three types of gratings of \cha\ and \xmm, namely the LETG, the HETG, and the RGS, cover wavelength ranges that cannot be measured with, e.g., {\it XRISM}. Losing them before a mission like {\it NewAthena} is available will critically prevent us from following important physical phenomena, from which we learn key facts of stellar, atomic physics, and gas dynamics. In particular, the LETG with the HRC camera will be absolutely unique for many years to come.

\cha\ and \xmm\ have strengthened the infrastructure of X-ray astronomy, laying the groundwork for subsequent and future missions. The synergy of ground-based and space-borne observations and surveys, conducted now and planned for the future, like  {\it XRISM, EINSTEIN Probe, eXTP, NewAthena, THESEUS, CTA, LISA, VERA RUBIN LSST, E-ELT, SKA}, and others, will provide key insights for new accomplishments in the AWD science, addressing current and future challenges.

\vspace{6pt}
\authorcontributions{\c{S}.B. is the main author responsible for conceptualization and organization of this review paper, together with the
writing of the abstract, main text, and the sections Introduction, Cataclysmic Variables, AM CVn Systems, and Final Remarks. M.O. is responsible for writing and organization of the section about Novae in Outburst. G.J.M.L. is responsible for the writing and organization of the section about WD Symbiotics. All authors have edited the main text for clarity. All authors have read and agreed to the published version of the manuscript.}

\funding{This { review } received no external funding.}


\acknowledgments{The authors thank J. Osborne, D. de Martino, and A. Schwope for comments and/or useful discussions on the manuscript. GJML is member of the CIC-CONICET (Argentina).}

\conflictsofinterest{The authors declare no conflicts of interest. }




\begin{adjustwidth}{-\extralength}{0cm}

\reftitle{References}


\PublishersNote{}
\end{adjustwidth}

\end{document}